\documentclass[manuscript]{acmart} 
\setcitestyle{numbers,sort&compress}
\usepackage{natbib} 

\acmConference{Equity and Access in Algorithms, Mechanisms, and Optimization}{October 5--9}{Virtual}

\title{Algorithmic Auditing and Social Justice: Lessons from the History of Audit Studies}

\author{Briana Vecchione}
\affiliation{%
\institution{Cornell University}
\city{New York}
\country{USA}}
\email{bv225@cornell.edu}

\author{Solon Barocas}
\affiliation{%
\institution{Microsoft Research and Cornell University}
\city{New York}
\country{USA}}
\email{solon@microsoft.com}

\author{Karen Levy}
\affiliation{%
\institution{Cornell University}
\city{Ithaca}
\country{USA}}
\email{karen.levy@cornell.edu}

\title{Algorithmic Auditing and Social Justice: Lessons from the History of Audit Studies}

\copyrightyear{2021}
\acmYear{2021}
\setcopyright{rightsretained}
\acmConference[EAAMO '21]{Equity and Access in Algorithms, Mechanisms, and Optimization}{October 5--9, 2021}{--, NY, USA}
\acmBooktitle{Equity and Access in Algorithms, Mechanisms, and Optimization (EAAMO '21), October 5--9, 2021, --, NY, USA}
\acmDOI{10.1145/3465416.3483294}
\acmISBN{978-1-4503-8553-4/21/10}

\begin{document}

\begin{abstract}

``Algorithmic audits” have been embraced as tools to investigate the functioning and consequences of sociotechnical systems. Though the term is used somewhat loosely in the algorithmic context and encompasses a variety of methods, it maintains a close connection to audit studies in the social sciences---which have, for decades, used experimental methods to measure the prevalence of discrimination across domains like housing and employment. In the social sciences, audit studies originated in a strong tradition of social justice and participatory action, often involving collaboration between researchers and communities; but scholars have argued that, over time, social science audits have become somewhat distanced from these original goals and priorities. We draw from this history in order to highlight difficult tensions that have shaped the development of social science audits, and to assess their implications in the context of algorithmic auditing. In doing so, we put forth considerations to assist in the development of robust and engaged assessments of sociotechnical systems that draw from auditing's roots in racial equity and social justice.

\end{abstract}

\maketitle

\renewcommand{\shortauthors}{Vecchione et al.}

\section{Introduction}

The past decade has been marked by an increasing push to subject algorithms to audit \cite{bandy2021problematic, kroll2016accountable, mittelstadt2016automation,  waldman2020privacy}. The call for such audits reflects a growing sense that algorithms play an important, yet opaque, role in the decisions that shape people’s life chances---as well as a recognition that audits have been uniquely helpful in advancing our understanding of the concrete consequences of algorithms in the wild and in assessing their likely impacts. 

As audits have proliferated, however, the meaning of the term has become ambiguous, making it hard to pin down what audits actually entail and what they aim to deliver \cite{brown2021algorithm, Ng_2021, Sloane_2021}. It is common for \textit{any} empirical investigation of an algorithm to be deemed an audit, despite the fact that such studies may involve very different measurement techniques and may focus on very different research questions. This confusion is understandable: colloquial use of the term audit can refer to a broad range of activities---including legal investigations by government tax authorities, compliance-oriented inspections by accounting firms, and the like---and the term is taken up in different ways in various disciplines (e.g., anthropology \cite{strathern2000audit} and public management \cite{reichborn2017performance}).

Recent work has attempted to give a more precise description of what algorithmic audits could---and should---entail, offering important methodological recommendations and raising challenging questions \cite{raji2020saving, raji2020closing, sloane2021silicon}. This paper aims to complement these efforts by drawing lessons from the history of audit studies, which originate in the social sciences. 

Starting in the 1970s, social scientists, advocates, and community organizers developed empirical methods to detect and measure the degree of discrimination in domains like housing. These initial efforts were driven by explicit concerns with racial equity and social justice, developed with direct participation of the affected communities and oriented around accountability and reform. Audit studies also represented an important methodological innovation, as they involved having real people go through the actual process of seeking housing, with the aim of measuring the degree to which characteristics like race influenced their treatment. Audit studies in the social sciences are a paradigmatic example of a “field experiment”---that is, a controlled experiment conducted not inside a lab, but instead out in the real world, with the goal of observing how actual decision makers behave. Such methods have been heralded as particularly effective techniques for uncovering the degree of discrimination that different groups face across various domains, especially as more overt signals of racial prejudice have declined \cite{pager2007use}. A particular version of the audit study---the “correspondence study,” in which researchers submit, for example, resumes to employers that systematically vary signals of gender or race---is now commonly viewed as the gold standard for empirical investigation of discrimination \cite{bertrand2017field}. 

While the methodological rigor of audit studies is one of their primary advantages, the merits of the technique have also been subject to debate within social science. As Cherry and Bendick point out in their excellent historical overview of audit studies, from which we will draw extensively: the “single-minded pursuit of rigor [in latter-day audit studies] risks sacrificing other considerations historically associated with auditing’s unique contributions to both society and science” \cite{cherry2018making}. In particular, scholars observe that field experiments to detect discrimination have become disembedded from communities subject to discrimination, and thus disconnected from their original social justice aims. 

These points of debate within the social sciences are particularly instructive for those conducting algorithmic audits. Our paper thus proceeds accordingly. After providing a brief history of audit studies in the social sciences, we describe the many ways in which algorithmic auditing is indebted to this prior work, while also frequently departing from it. We then discuss how the history of audits in the social sciences highlights a number of difficult tensions that those conducting algorithmic audits would do well to consider---and how algorithmic audits might be designed and conducted in a way that accounts for the racial equity and social justice roots of auditing in social science. 

\section{The Social Justice Origins of Social Science Auditing}

Social science auditing originated in activist research. In the 1940s and 1950s, civic organizations partnered with researchers in order to better assess the state of race relations in the United States \cite{cherry2018making, gaddis2018introduction}. Since civil rights were not yet statutorily protected, these groups had to rely on ``persuasion and cooperation” to effect change; the currency of this strategy was \textit{information} about the incidence of discrimination, which audit studies could provide. Community groups marshalled audit studies to raise awareness of problems---including, in some cases, community groups meeting with audited entities after-the-fact to discuss what the study had revealed about their specific behavior \cite{cherry2018making}.

For researchers, these collaborations were a key example of the participatory action research (PAR) paradigm. PAR methods involve collaboratively and reflexively engaging with research participants, and treating them as more than merely informants in a researcher-led study. Instead, researchers and practitioners co-design and co-execute projects from the ground up, with dual goals of both ``advancing scientific knowledge [and] achieving practical objectives” \cite{whyte1989advancing}. Participatory action emphasizes the importance of conducting research \textit{with} participants, not on or for them \cite{mcintyre2007participatory}. 

Methods that arose during this time included techniques like community self-surveys, in which community members themselves collected data via in-person and/or telephone interviews. Communities themselves organized the studies, with support and training from researchers. A key priority of community-led studies was the experience of participation in research itself. Engaging in the research process was understood to support community autonomy, increase trust in the results, and spur reform. As the foreword to the 1951 manual \textit{How to Conduct a Community Self-Survey of Civil Rights} explains: ``Knowledge self-obtained seems more authentic than second-hand knowledge. It reflects both an inner sense of urgency and a dependable view of social reality. Thus it is pointed toward effective social action. Since the self-survey leads to full-bodied participation on the part of the citizen it is a valuable tool of modern democracy”  \cite{wormser1950conduct}.

Later, audit studies became an important tool for enforcement of civil rights statutes. This was most notable in housing, where large-scale in-person audits in the 1970s---sponsored by the Department of Housing and Urban Development, and executed in partnership with community groups and local researchers---demonstrated the prevalence of discrimination against Black prospective tenants and homebuyers across the country, and influenced federal policy \cite{cherry2018making, yinger1998evidence}. 

Over time, however, enthusiasm for community/researcher partnerships to study discrimination waned. Participatory action perspectives in general became less common within the academy, as partnerships were resource-intensive and poorly incentivized. Concomitantly, more sophisticated quantitative modeling gained traction across the social sciences, creating expectations for greater statistical and methodological rigor in academic work \cite{cherry2018making}. In in-person audits, even when testers are carefully matched---when the researcher takes care to control for factors like age, physical appearance, and the like (that is, everything but the variable being tested)---unobservable differences and experimenter effects may still confound the study and are difficult to measure \cite{pager2007use}. And by the 2000s, many employment and housing application processes occurred online rather than in-person, making them less amenable to in-person audits \cite{gaddis2018introduction}.

These dynamics have contributed to a trend away from in-person audits and toward correspondence studies. Correspondence studies afford researchers great control over the variables presented, and can be conducted at much broader scale than in-person audits, including across many geographic contexts \cite{besbris2018geography, pedulla2018emerging}. In these studies, researchers rely on fictitious correspondence by hypothetical individuals online or by mail, systematically varying the characteristics of ``applicants” to test for discrimination. In one well-known correspondence audit by Bertrand and Mullainathan \cite{bertrand2004emily}, researchers applied to over 1,300 newspaper job advertisements via fax and mail, sending four systematically varied resumes to each employer. The study's findings were striking: all else equal, Black applicants were about 50\% less likely than white applicants to receive a callback. The study was also remarkable due to its unprecedented scale, demonstrating that a large-scale correspondence audit could be done by only a few researchers. Correspondence studies, of course, face their own methodological limitations---a notable difficulty is how effectively researchers can signal the characteristic of interest (and only the characteristic of interest) using applicant names or other indicators \cite{gaddis2017black, crabtree2018last}---but their comparative advantages have led them to dominate audit studies in the social sciences, and to be considered a ``significant methodological advance’’ over in-person tests \cite{bertrand2017field, guryan2013taste}. Correspondence studies in the social sciences have been used to examine discrimination involving a host of personal characteristics other than race and gender, including employment history \cite{pedulla2016penalized}, LGBT status \cite{mishel2016discrimination}, immigration status \cite{gell2018s}, and many other characteristics \cite{gaddis2018introduction}.

\section{Algorithmic Auditing}

In some ways, algorithmic audits have much in common with audit studies in the social sciences, suggesting that lessons from the latter should easily carry over to the former. Yet recent work has shown that the goals motivating activities deemed algorithmic audits---and the methods used to conduct them---can be surprisingly capacious \cite{bandy2021problematic} and often differ from those in the social sciences. In the developing academic and policy discourse---and increasingly in colloquial language---the term algorithmic audit has come to refer to almost any kind of empirical study of algorithms, even though audit studies within the social sciences refer specifically to field experiments designed to measure the extent of discrimination within the domain. This degree of variation in algorithmic audits has created some confusion about what the term algorithmic auditing means and what qualifies an activity as an algorithmic audit. 
Understanding variations in both the goals and methods of algorithmic audits can help to ensure that the lessons we draw from the social sciences are well-matched to the particular type of algorithmic audits for which they are relevant.

While some algorithmic audits are focused on establishing whether an algorithm considers legally proscribed factors or whether an algorithmic process is sensitive to changes in these factors \cite{ali2019discrimination, eslami2017careful, hannak2017bias, imana2021auditing, markup2021mortgage, wilson2021building}, other algorithmic audits have different goals. For example, some algorithmic audits attempt to uncover violations of procedural regularity---that is, when an algorithm fails to function as promised and in a consistent manner, but without specific emphasis on discrimination \cite{datta2015automated, kroll2016accountable, robertson2021can, saxena2021framework}. Others are motivated by a concern with transparency, aiming to uncovering how an algorithm works---with the term auditing often meaning something like reverse engineering \cite{AdaLovelace, adler2018auditing, diakopoulos2015algorithmic}. Even traditional evaluations of algorithms’ performance in terms of accuracy, precision, and recall \cite{gunawardana2009survey} are sometimes described as audits; more recent efforts have begun to disaggregate these results by group, revealing when algorithms perform less well for some groups than others \cite{angwinmachinebias, barocas2021designing, buolamwini2018gender}---a problem often described as bias, but different in nature than the discriminatory treatment examined by social science audits. Still others focus on completely different notions of bias, like the balance of political views in algorithmically curated content \cite{robertson2018auditing}.

While audit studies in the social sciences are by definition experiments (in which the researcher directly manipulates the variable of interest, and then measures outcomes), the wide range of studies described as algorithmic audits can involve purely observational studies of algorithmic outcomes (in which the researcher measures outcomes without manipulation) as well as direct inspection of the algorithms themselves. Algorithmic audits that take an experimental approach often rely on so-called ``sock puppets,” where computer programs are used to impersonate users of a platform \cite{asplund2020auditing, sandvig2014auditing}. This is similar to the approach taken in correspondence studies in the social sciences, but with a greater degree of automation. Like correspondence studies, sock-puppet-based algorithmic audits allow researchers to carefully vary inputs, furnishing greater statistical rigor. Researchers can also enlist volunteers to participate in an audit---a practice Sandvig et al. \cite{sandvig2014auditing} describe as a ``noninvasive user audit.” In this case, real users  disclose information about their interactions with a platform (e.g., search queries and results), with the goal of allowing researchers to learn something about the platform’s algorithm, despite the lack of experimental control. A recent example is The Markup’s ``citizen browser project,” where a panel of individuals installed a custom browser on their computers that tracks data about their Facebook and YouTube accounts, hopefully leading to insights about how each platform’s algorithms operate \cite{markup2020}.

In what follows, we aim to draw lessons from the social sciences that are most relevant to those who look to audits as a way to uncover and address discrimination in algorithms specifically. In narrowing our focus to audits concerned with discrimination, we do not intend to suggest that audits cannot or should not play a role in evaluating algorithmic systems with other normative concerns in mind. Nor do we mean to suggest that insights from the history of audit studies in the social sciences are irrelevant to algorithmic audits with a different focus. However, it is difficult to discuss the relative merits of different approaches to audits when the purpose of such audits is left under-specified. We therefore focus the rest of our discussion on audits that aim to uncover and root out discrimination.

\section{The Roots of Social Science Auditing and the Future of Algorithmic Auditing}

The history of auditing in the social sciences can help us chart a path forward for algorithmic auditing. Below, we describe four key considerations, each of which draws from tensions within social science regarding the aims and capabilities of audit studies, and describe their implications in the context of algorithmic audits. Our goal is to draw from the past in order to help us better reflect on the future: we should learn from the lessons of social science to intentionally consider what role we want algorithmic audits to play in advancing social justice and how to go about designing and conducting them. In so doing, we again note the debt that our analysis owes to Cherry and Bendick’s insightful critique of audits in the social sciences, which emphasizes their roots in the activist scholarship tradition \cite{cherry2018making}.

\subsection{Beyond discrete moments of decision making} Social science audit studies can make claims only about discrete \textit{moments} at which bias can emerge---and be readily measured---in a process \cite{barocas2017fairness}. For example, many correspondence studies in the hiring context zero in on the decision to invite an applicant for an interview based on an evaluation of their resume; earlier and later stages of hiring processes, and the biases that may emerge in them, are not captured. In an analysis by Quillian et al. \cite{quillian2020evidence} of over 100 audit studies of racial discrimination in the employment process, only 13 considered the ultimate employment outcome (a job offer), choosing instead to treat the ``callback''---an invitation to interview---as a proxy for the outcome of interest. Despite this, they show that substantial additional discrimination occurs \textit{after} the callback stage, and is missed by most studies.

There are good reasons for such narrow focus in social science auditing. As Cherry and Bendick describe, social scientists are drawn to correspondence studies over in-person audits for many reasons, including their amenability to making rigorous statistical claims, high degree of researcher control, and ability to scale. These methods naturally lend themselves to examining only certain stages of social processes in which people can be evaluated based on indicators that can be ascertained through documents. Doing so can also help to eliminate noise in the data, and can focus findings on specific mechanisms through which bias proliferates. Finally, research ethics may require that researchers restrict field experiments to parts of a process that are unlikely to cause significant harm to research participants and others (particularly in light of the deception that audit studies require); the cost to an employer of evaluating an extra resume is much lower than that of interviewing a live tester, who may be displacing a genuine job candidate.

Therefore, while correspondence audits can be rigorously controlled and demonstrate statistically significant effects, they can offer only a ``thin slice” view of where in a process discrimination can occur. Other social scientific methods can complement audits to shed light on other stages in a process. For example, Rivera’s ethnographic examination of elite firm hiring practices delineates nine stages of hiring, from recruitment to deliberation \cite{rivera2016pedigree}. Rivera demonstrates that privileged applicants are advantaged across the board---not only by resume indicators (e.g., university attended) but by class markers that emerge elsewhere in the process, like in in-person interviews (e.g., cultural homogeneity and interaction style). Because some of these stages are less amenable to audit methods, audits that focus only on resume screening necessarily underestimate the true degree of discrimination job applicants face in a hiring process. 

Similar concerns arise in the context of algorithmic auditing. \citet{selbst2019fairness} caution researchers who study fairness in machine learning against the \textit{framing trap}---the ``failure to model the entire system over which a social criterion, such as fairness, will be enforced.” They encourage researchers to analyze not only how bias may arise in a machine learning model in isolation, but how it may arise \textit{sociotechnically}, when humans and institutions interact with the model in the social world. For example, a judge receiving recommendations from a risk assessment tool \cite{stevenson2018assessing} or a hiring manager obtaining a ``fair” list of screened job applicants may make decisions in concert with the technical tool that simply relocates bias to a new, opaque stage of the process. At the very least, algorithmic audits that focus solely on the technical components of a sociotechnical system should be understood to be subject to this limitation. But often, the distinction is blurred; a process may be trumpeted as ``fair” with respect to its technical components, without due attention to the caveat that this assessment applies at best to only particular stages of a process \cite{bogen2018help, raghavan2020mitigating}.

We might also attempt more ``end-to-end” audits that encompass humans and institutions in interaction with algorithmic systems, accounting for operational and social aspects of use alongside technical considerations \cite{introna2010facial}. Ample evidence suggests that best-case assumptions about how an algorithm is used---which might be the basis for audits proclaiming they have met some fairness or quality criterion---may not hold empirically. A clear example emerged when the ACLU demonstrated that Amazon’s face recognition product mistakenly matched 28 Members of Congress to mugshots: Amazon countered that the ACLU had intentionally misrepresented the product by failing to use its recommended 99\% confidence threshold for public safety applications \cite{gizmodoFaceRec, acluFaceRec}. The ACLU countered that it had used an 80\% confidence threshold which was the \textit{default setting} for the product, and therefore likely to reflect actual use, no matter what Amazon’s manual recommended. (Research demonstrates that police departments \textit{do} routinely engage in all kinds of less-than-ideal practices around face recognition algorithms---including using both composite sketches and celebrity lookalike photos as input data \cite{garbageInGarbageOut}). On the other ``end,” Stevenson and Doleac demonstrate how human judges inconsistently \textit{follow} the sentencing recommendations proffered by risk assessment algorithms in the criminal justice context, potentially reintroducing biases and further skewing outcomes by race and age \cite{IZAalgRiskAssessments}. The key point is that audits might be most meaningfully deployed to assess bias in \textit{actual, not optimal}, conditions of end-to-end use.

More broadly, both social scientific and algorithmic auditing necessarily underestimate both the incidence and impact of discrimination in the social world. In addition to the ``thin-slice” problem we have detailed, it is important to recognize that many discriminatory processes cannot be examined via audits at \textit{any} stage. For example, jobs that are filled through social networks rather than through job ads cannot be readily audited \cite{baldassarri2017field}, despite being both frequent routes to employment \cite{granovetter1995getting, trimble2011role} and likely to propagate discriminatory outcomes due to homophily \cite{boyd2014networked, mcpherson2001birds}. Blue-collar jobs more often require in-person application than white-collar jobs, and thus may be less readily auditable using correspondence studies \cite{pager2007use}. Perhaps most importantly, audits typically assess \textit{episodic} incidents of discrimination---one job search, one credit application---not discrimination that accrues \textit{cumulatively}. But as Small and Pager note, discrimination causes harm “not just at critical junctures but also over the slow, lifelong buildup of its everyday sting” \cite{small2020sociological}. While we should appreciate what can be learned from audit studies, we should also appreciate what can’t be.

\subsection{The dangers of adopting and abandoning experimental controls} 

Audit studies aim to measure the degree to which differences in the perceived gender or race of job applicants, for example, affect the way that employers treat candidates who are otherwise identical. In this respect, audit studies are an attempt to detect what the law deems ``disparate treatment”: decision making that takes legally proscribed variables like gender or race into account. The fundamental premise of audit studies is simple, but powerful: if decision makers treat candidates differently who are the same except for their gender or race, then the decision must be taking this feature into account, as nothing else could explain the difference. Tightly controlling for other factors that might influence the decision thus allows researchers to conclude that the decision maker's awareness of applicants' gender or race must be influencing the outcome. This explains the enormous effort that researchers invest in training in-person testers, aiming to ensure that testers present equivalent background stories, answer questions in the same way, and otherwise comport themselves in a manner that makes them as similar as possible except for their gender or race (more on this challenge in a moment) \cite{pager2007use}. It likewise explains the allure of correspondence studies, which allow researchers to avoid this challenge altogether, as the testing takes place via paper resumes over which researchers have complete control.

Unfortunately, these methods do not always translate well to algorithmic decision making because many algorithms used in high-stakes domains (like hiring) do not include gender or race as direct inputs to the decision.\footnote{This is not to say that potential---and sometimes quite obvious---proxies for gender and race might not remain part of the process. In the well-cited Reuters story about Amazon’s abandoned AI recruitment tool, the tool was found to penalize job candidates whose resumes included language like “women’s chess club captain,” giving away applicants’ gender, even if it was not an explicit feature \cite{Dastin_2018}. To our knowledge, however, no algorithmic audits have tried to test experimentally whether algorithms differ in their behavior when such features are varied while everything else is held constant. Experiments along these lines would be very similar to those that have been conducted in the social sciences over the past decade, which, beyond using names as a signal of gender or race, also manipulate other aspects of job applicants’ resume like professional associations and personal interests that encourage employers to draw similar inferences \cite{pager2007use}.} While there are a few instances of algorithmic audits that have tried to establish whether algorithms are indeed engaged in disparate treatment \cite{datta2015automated, sweeney2013discrimination}---and, in fact, find that they are---the vast majority aim to measure what the law calls ``disparate impact": differences in the outcomes experienced by difference groups even when the decision making process only relies on seemingly benign variables. This approach abandons the overriding focus on experimental control that is characteristic of the more recent audit studies in the social sciences and instead aims to measure how different groups fare when subject to algorithmic decision making, \textit{given natural differences between groups in the variables that the algorithm takes into account}. In many respects, this is a return to the approach characteristic of earlier audit studies.

Changing attitudes within the social sciences about the relative merits of each approach offer helpful lessons for those undertaking algorithmic audits. Carefully controlled audit studies offer a degree of methodological rigor that makes it much easier to attribute different outcomes directly to the variable of interest. In the absence of careful controls, advocates argue, differences in outcome (e.g., the callback rate) might be explained by differences in other covariates (e.g., formal qualifications, education, work experience, etc.), the distribution of which might differ across groups. It is far more challenging to make an argument that decision makers have discriminated on the basis of gender or race when there are other potential explanations for the disparity in outcomes. But the criticisms of tightly controlled experimental approaches in the social sciences also reveal the danger of treating the process of establishing discrimination as a matter of showing that certain outcomes would have been the same, but for a difference only in some discrete marker of gender or race \cite{sen2016race}. As Hu and Kohler-Hausmann argue, gender and race are not isolated attributes disconnected from all other facts about a person---attributes that can be varied experimentally without affecting other factors traditionally subject to tight control \cite{hu2020s}. What it \textit{means} to be a woman or be a Black person is not determined by which gender or race a person happens to tick on a form---or have ticked for them. These are complex social constructs that implicate a profound range of details about a person and the way others interact with them \cite{bonilla2008toward}. When researchers control for everything but some discrete marker of gender or race (e.g., a name), they overlook how gendered and racialized identity necessarily implicates the many other factors that are subject to careful experimental control \cite{zuberi2008white}.

Many algorithmic audits have---perhaps unwittingly---avoided some of this tension by focusing instead on differences in the accuracy of these decisions. This approach seems to rest on an intuitive normative belief that while people from different groups may not be entitled to the same outcomes, given possible differences in the qualities possessed by members of these groups, they are all entitled to equally accurate assessment. This is quite a departure from the focus of audit studies in the social sciences (both the early and more recent variants), which have focused less on evaluating the accuracy of decision making, in part due to methodological limitations (e.g., researchers would need to know how well a job applicant would have done had they been hired) and in part due to the way that they conceptualize discrimination (i.e., as differences in treatment, given differences in overt markers of gender or race.\footnote{Whether the difference in accuracy is attributable to differences in gender or race or to other features that happen to correlate with gender or race is irrelevant to this analysis, as the goal is to simply show that an algorithm's accuracy varies between groups given natural differences in their covariates.} Along the same lines, some algorithmic audits seek to document that a decision-making process produces disparities in outcomes for different groups without controlling for any differences between these groups. Certain audits treat these differences in outcomes as inherently suspect---and are thus vulnerable to criticisms about a lack of appropriate control. Others, however, attempt to identify the factors in the decision-making process that contribute to the disparity in outcome so that their legitimacy as decision criteria might be critically assessed. Rather than testing whether specific pre-determined factors (e.g., gender or race) matter to a decision, as is the approach in audit studies, these algorithmic audits asks which factors---among all those under consideration---explain the difference in outcomes and whether these are problematic in some way. 

Each of these approaches to algorithmic audits do not try to manipulate a specific marker of gender or race while holding other features constant; instead, they take the natural differences in the distribution of these features by gender or race as a given and then evaluate how the algorithm handles these cases, looking for systematic differences in accuracy or outcomes. As such, claims about the degree to which an algorithm results in differences in accuracy or outcomes depend entirely on the distribution of features in the populations that have been used to perform the evaluation \cite{barocas2021designing}. In practice, many algorithmic audits are performed with the population that is in the training data, but this population may not reflect the population that will be subject to the algorithm. What appears to be no notable difference in accuracy or outcomes between groups in the training data might not hold when the model encounters future populations from each group with a distribution of features that are quite different. It is not possible to make general claims about a model’s differential performance, its potential to cause disparate impact, or the cause of these disparities without making certain assumptions about the likelihood that the distribution of features in the evaluation dataset will match the distribution in the target domain \cite{raghavan2019challenges}. And yet, these assumptions are rarely disclosed, including in such high-profile cases like the recent audit of HireVue’s automated assessment tool conducted by O'Neil Risk Consulting \& Algorithmic Auditing (ORCAA) \cite{orcaa}. ORCAA’s report, which finds that there was no ``bias” in the assessment, does not offer any details about the dataset used to perform the evaluation. Failing to disclose this information might be a carry-over from carefully controlled audit studies, where it was easy to infer what data had been used because the goal was to use the same data (e.g., resume) in all cases while only varying gender or race. This is \textit{not} the case with most algorithmic audits and thus much more detailed disclosure is necessary to be able to interpret the results.

\subsection{Forms of knowledge and forms of evidence}

An important consideration in the design of an audit study is \textit{what type of knowledge} it seeks to produce. Does an audit study seek to create generalized knowledge about the ``state of the world’’---for example, an average or aggregate measure of the degree of discrimination present in a labor market? Or does it seek to create specific knowledge about the practices of a \textit{particular} employer, landlord, or credit provider? Both forms of knowledge can be valuable. Social scientific audit studies have tended to favor the former---unsurprisingly so, given social science’s emphasis on describing the social world. But we might imagine that communities affected by discrimination may be less interested in such general knowledge, and more interested in knowledge that supports action targeted at specific discriminating institutions.

These different goals may lead to different audit study designs. For example, many contemporary audit studies are \textit{unpaired}---that is, treatment and control testers are sent to \textit{different} randomized recipients, and collective outcomes compared statistically in order to assess the amount of discrimination experienced in aggregate \cite{cherry2018making}. In contrast, in a \textit{paired} audit, each recipient evaluates two applicants who are matched on all characteristics except for the variable of interest. Unpaired audits may be more efficient for researchers: in some cases they can be more statistically powerful, and can reduce the risk that entities become suspicious that they are being subjected to an audit \cite{, vuolo2016statistical, vuolo2018match}. But unpaired audits do not allow for identification of \textit{specific} discriminating decision makers, since no single entity assesses multiple testers. As Cherry and Bendick describe, unpaired audits report ``villainy without villains’’---they ``document an abstract evil attributable only to the overall population from which the audit sample was drawn” \cite{cherry2018making}.\footnote{That said, even paired audits may not be able to demonstrate statistically significant discrimination if each recipient only receives (say) one pair of resumes, but they at least provide one pair of comparative data points for each audited entity.} But because algorithmic audits \textit{do} often focus on evaluation of a specific, identifiable system, they may be able to avoid the ``villainy without villains'' problem, and may be put to use for different forms of social activism, as early social scientific audits did.\footnote{Of course, there may also be virtues to using algorithmic audits to create generalized knowledge, perhaps by auditing a \textit{set} of systems in comparison to one another. For example, a recent examination by Rieke et al. \cite{riekeupturn} of hiring platforms used by 15 large hourly employers is able to make general claims about common obstacles facing job-seekers by virtue of its comparative research design. One could easily imagine an algorithmic audit study that similarly assesses discrimination across multiple platforms.}

A final note on this point involves the form and communication of empirical findings from audit studies, and the degree to which such knowledge can support concrete social change. In-person audits have the virtue of narrative depth: live testers’ vivid descriptions of their actual experiences being subjected to biased treatment can be extremely powerful for capturing public attention and galvanizing reforms \cite{cherry2018making, bendick2012developing, pedulla2018emerging}. In-person audits thus can produce ``both stories and statistics” \cite{cherry2018making}, each of which has its own persuasive authority to effect change. Paper-based correspondence studies are less likely to produce compelling qualitative accounts since no real person is subjected to biased treatment.

In the algorithmic realm, both stories and statistics have a role to play. Audit researchers have compellingly integrated vivid qualitative accounts---for example, Joy Buolamwini’s experience of being invisible to face recognition software unless she wore a white mask \cite{NYT}---alongside rigorous analysis quantifying the extent of the problem. Other researchers have provided compelling narratives to communicate audit findings, like the aforementioned ACLU demonstration that Amazon’s face recognition product erroneously matched 28 Members of Congress (including, notably, civil rights hero Rep. John Lewis) to mugshots \cite{acluFaceRec}, or ProPublica’s famed investigation of the COMPAS risk assessment algorithm, which paired statistical analysis with the stories of individuals subjected to it \cite{angwinmachinebias}. Efforts to integrate quantitative findings alongside compelling narratives of individual cases are promising ways of achieving non-academic policy impacts that can serve community goals more effectively than either approach in isolation.

\subsection{Ensuring meaningful community alliances}

Beyond the creation of different forms of knowledge and evidence, audit studies may differ in terms of their type and degree of engagement with community groups. As Cherry and Bendick document, in early audits, partnerships with community groups were considered ``as integral an objective of the activity as were published reports'' \cite{cherry2018making}, drawing from a participatory action research tradition. Over time, such collaboration waned, as social scientists began to focus more on abstract modeling and correspondence studies, which lacked roles for human testers from community groups.

In the algorithmic sphere, scholars have remarked on the degree to which the framing of fairness research may be disconnected from the actual priorities of affected communities, and the tendency to turn the lived experience of discrimination into an abstract intellectual exercise \cite{d2020data, hoffmann2019fairness}. Therefore, we should ask: what might meaningful community engagement look like in algorithmic auditing? To what degree could researchers collaborate with communities in designing their approaches, and what forms might such collaboration take?

One potentially promising avenue is the development of crowdsourced or ``collective” community audits. Crowdsourced audits might involve soliciting participation from willing volunteers who share their data with researchers. A good example is the coordinated use of the GDPR’s subject access rights, which allow any individual to obtain a record of their personal data and information about how their data have been  used in automated decision making systems. Individuals who trigger data access requests about themselves may then ``donate” their data to researchers, who use these data in aggregate to audit systems for discrimination and reverse-engineer their functioning \cite{ausloos2020researching}. This collective approach can be understood as a means of rebalancing information asymmetries in the service of social justice \cite{mahieu2020recognising}. Another approach borrows from the successful model of ``bug bounty” programs, which crowdsource security testing by offering researchers financial incentives for identifying security vulnerabilities. Algorithmic bug bounties may function similarly by providing infrastructure for users to report perceived algorithmic biases in the services they use \cite{TwitterBugBounty, vice, TwitterBountyInsights}.

``Organic” audits are a related development: spontaneous, community-led efforts that arise as a result of perceived bias in a public-facing algorithm, often disseminated through social media. For example, in 2019, entrepreneur David Heinemeier Hansson noted on Twitter \cite{dhh} that his Apple Card credit limit was 20 times that of his wife---despite them having access to all the same assets, and despite customer service’s assurances that the algorithm was fair. The Twitter thread went viral with many others (including Steve Wozniak!) noting that they’d encountered similar issues, and communicating the broad strokes of their own application characteristics and outputs. Of course, this ``audit” lacked systematic sampling, experimental control, or statistical rigor, and as such lacked the hallmarks of a scientific study, but it drew enough attention to the problem that the issue was covered in prominent media and sparked an investigation from New York’s State Department of Financial Services \cite{washPost} (which eventually cleared Apple of any wrongdoing \cite{Nasiripour_Farrell_2021}). A similar dynamic unfolded when a student posted on Twitter that Zoom routinely ``removed the head” of a Black colleague when that colleague used a virtual background, suggesting that Zoom’s algorithms failed to adequately detect darker skin tones \cite{NPR}. When the student posted various images to Twitter to illustrate the problem, Twitter’s image cropping algorithms repeatedly previewed only those parts of images with white faces---suggesting that Twitter’s algorithms \textit{also} suffered from algorithmic bias in determining what part of an image was most salient to preview (despite the fact that Twitter had already conducted an internal bias audit on its system). This was quickly followed by many others posting their own sets of images with similar results. The event prompted Twitter to announce it was revisiting its image cropping algorithms and changing its interface to allow users more control over image previews \cite{twitterTransparency,yee2021image}.

Community-based auditing techniques are advantageous in that they can surface potential biases and test cases that internal auditors might overlook. They also have the important advantage of fostering community awareness and mobilization much more than a researcher-led or internal audit likely would---and concomitantly, greater public accountability. But these techniques also may be less rigorously controlled and systematically sampled than a centralized, researcher-led audit would be, and subject to greater data quality concerns from less-well-trained volunteers. These are important trade-offs---and quite similar to those confronted by other distributed data collection and ``citizen science” efforts, which have developed a number of best practices to address sampling biases and achieve data quality \cite{salganik2019bit, sullivan2014ebird}. Another approach is to combine community- and researcher-led approaches collaboratively \cite{sandvig2014auditing} by using community results to identify areas for researchers to subsequently probe systematically (akin to qualitative theory-building and quantitative theory-testing in social science research \cite{eisenhardt2007theory}). More centralized researcher control---like the collection of data from a representative panel---can address concerns about unsystematic selection of test cases, but seems less likely to result in the meaningful community engagement and capacity-building that characterized early participatory audits.

\section{Conclusion}
In this work, we have highlighted aspects of the history of audit studies in the social sciences---and ongoing debates about their design and implementation---that are useful for the development of algorithmic auditing. The history suggests that there is significant valuable knowledge that can be generated by carefully controlled experiments conducted in contexts marked by discrimination. But statistical rigor comes at a price. Earlier orientations to auditing had different desirable features that have been diminished over time---particularly with respect to community engagement.

As auditing develops as a research method in the algorithmic context, we would do well to heed this history, and to acknowledge the trade-offs that different approaches entail. How expansive should we be in defining discrimination, and what \textit{can't} we measure using audit methods? What are the advantages of documenting the experiences of actual people, even if doing so diminishes our ability to  make rigorous causal claims? What forms of knowledge and evidence should we be trying to create to support affected groups, and how should we engage people meaningfully in research about discrimination? The answers to these questions are not straightforward---but considering lessons from the past can help us reflect critically on the role of algorithmic auditing in fostering accountability and change.

\section{Acknowledgments}

We thank the John D. and Catherine T. MacArthur Foundation for support and Cornell University's Artificial Intelligence, Policy, and Practice (AIPP) Initiative and David Pedulla for valuable feedback.

\raggedright
\bibliography{acmart}


\begin{thebibliography}{96}


\ifx \showCODEN    \undefined \def \showCODEN     #1{\unskip}     \fi
\ifx \showDOI      \undefined \def \showDOI       #1{#1}\fi
\ifx \showISBNx    \undefined \def \showISBNx     #1{\unskip}     \fi
\ifx \showISBNxiii \undefined \def \showISBNxiii  #1{\unskip}     \fi
\ifx \showISSN     \undefined \def \showISSN      #1{\unskip}     \fi
\ifx \showLCCN     \undefined \def \showLCCN      #1{\unskip}     \fi
\ifx \shownote     \undefined \def \shownote      #1{#1}          \fi
\ifx \showarticletitle \undefined \def \showarticletitle #1{#1}   \fi
\ifx \showURL      \undefined \def \showURL       {\relax}        \fi
\providecommand\bibfield[2]{#2}
\providecommand\bibinfo[2]{#2}
\providecommand\natexlab[1]{#1}
\providecommand\showeprint[2][]{arXiv:#2}

\bibitem[\protect\citeauthoryear{{Ada Lovelace Institute}}{{Ada Lovelace
  Institute}}{2020}]%
        {AdaLovelace}
\bibfield{author}{\bibinfo{person}{{Ada Lovelace Institute}}.}
  \bibinfo{year}{2020}\natexlab{}.
\newblock \bibinfo{title}{{Examining the Black Box}}.
\newblock
  \bibinfo{howpublished}{\url{https://www.adalovelaceinstitute.org/report/examining-the-black-box-tools-for-assessing-algorithmic-systems/}}.
\newblock


\bibitem[\protect\citeauthoryear{Adler, Falk, Friedler, Nix, Rybeck,
  Scheidegger, Smith, and Venkatasubramanian}{Adler et~al\mbox{.}}{2018}]%
        {adler2018auditing}
\bibfield{author}{\bibinfo{person}{Philip Adler}, \bibinfo{person}{Casey Falk},
  \bibinfo{person}{Sorelle~A Friedler}, \bibinfo{person}{Tionney Nix},
  \bibinfo{person}{Gabriel Rybeck}, \bibinfo{person}{Carlos Scheidegger},
  \bibinfo{person}{Brandon Smith}, {and} \bibinfo{person}{Suresh
  Venkatasubramanian}.} \bibinfo{year}{2018}\natexlab{}.
\newblock \showarticletitle{Auditing Black-Box Models for Indirect Influence}.
\newblock \bibinfo{journal}{\emph{Knowledge and Information Systems}}
  \bibinfo{volume}{54}, \bibinfo{number}{1} (\bibinfo{year}{2018}),
  \bibinfo{pages}{95--122}.
\newblock


\bibitem[\protect\citeauthoryear{Ali, Sapiezynski, Bogen, Korolova, Mislove,
  and Rieke}{Ali et~al\mbox{.}}{2019}]%
        {ali2019discrimination}
\bibfield{author}{\bibinfo{person}{Muhammad Ali}, \bibinfo{person}{Piotr
  Sapiezynski}, \bibinfo{person}{Miranda Bogen}, \bibinfo{person}{Aleksandra
  Korolova}, \bibinfo{person}{Alan Mislove}, {and} \bibinfo{person}{Aaron
  Rieke}.} \bibinfo{year}{2019}\natexlab{}.
\newblock \showarticletitle{Discrimination through Optimization: How Facebook's
  Ad Delivery can Lead to Biased Outcomes}.
\newblock \bibinfo{journal}{\emph{Proceedings of the ACM on Human-Computer
  Interaction}} \bibinfo{volume}{3}, \bibinfo{number}{CSCW}
  (\bibinfo{year}{2019}), \bibinfo{pages}{1--30}.
\newblock


\bibitem[\protect\citeauthoryear{Angwin, Larson, Mattu, and Kirchner}{Angwin
  et~al\mbox{.}}{2016}]%
        {angwinmachinebias}
\bibfield{author}{\bibinfo{person}{Julia Angwin}, \bibinfo{person}{Jeff
  Larson}, \bibinfo{person}{Surya Mattu}, {and} \bibinfo{person}{Lauren
  Kirchner}.} \bibinfo{year}{2016}\natexlab{}.
\newblock \bibinfo{title}{Machine Bias}.
\newblock
  \bibinfo{howpublished}{\url{https://www.propublica.org/article/machine-bias-risk-assessments-in-criminal-sentencing}}.
\newblock


\bibitem[\protect\citeauthoryear{Asplund, Eslami, Sundaram, Sandvig, and
  Karahalios}{Asplund et~al\mbox{.}}{2020}]%
        {asplund2020auditing}
\bibfield{author}{\bibinfo{person}{Joshua Asplund}, \bibinfo{person}{Motahhare
  Eslami}, \bibinfo{person}{Hari Sundaram}, \bibinfo{person}{Christian
  Sandvig}, {and} \bibinfo{person}{Karrie Karahalios}.}
  \bibinfo{year}{2020}\natexlab{}.
\newblock \showarticletitle{Auditing Race and Gender Discrimination in Online
  Housing Markets}. In \bibinfo{booktitle}{\emph{Proceedings of the
  International AAAI Conference on Web and Social Media}},
  Vol.~\bibinfo{volume}{14}. \bibinfo{publisher}{AAAI Press},
  \bibinfo{address}{Menlo Park, CA}, \bibinfo{pages}{24--35}.
\newblock


\bibitem[\protect\citeauthoryear{Auditing}{Auditing}{2020}]%
        {orcaa}
\bibfield{author}{\bibinfo{person}{O’Neil Risk Consulting \&~Algorithmic
  Auditing}.} \bibinfo{year}{2020}\natexlab{}.
\newblock \bibinfo{booktitle}{\emph{{Description of Algorithmic Audit:
  Pre-built Assessments}}}.
\newblock \bibinfo{type}{{T}echnical {R}eport}. \bibinfo{institution}{O’Neil
  Risk Consulting \& Algorithmic Auditing}.
\newblock


\bibitem[\protect\citeauthoryear{Ausloos and Veale}{Ausloos and Veale}{2021}]%
        {ausloos2020researching}
\bibfield{author}{\bibinfo{person}{Jef Ausloos} {and} \bibinfo{person}{Michael
  Veale}.} \bibinfo{year}{2021}\natexlab{}.
\newblock \bibinfo{title}{Researching with Data Rights}.
\newblock , \bibinfo{numpages}{136--157}~pages.
\newblock


\bibitem[\protect\citeauthoryear{Baldassarri and Abascal}{Baldassarri and
  Abascal}{2017}]%
        {baldassarri2017field}
\bibfield{author}{\bibinfo{person}{Delia Baldassarri} {and}
  \bibinfo{person}{Maria Abascal}.} \bibinfo{year}{2017}\natexlab{}.
\newblock \showarticletitle{Field Experiments across the Social Sciences}.
\newblock \bibinfo{journal}{\emph{Annual Review of Sociology}}
  \bibinfo{volume}{43} (\bibinfo{year}{2017}), \bibinfo{pages}{41--73}.
\newblock


\bibitem[\protect\citeauthoryear{Bandy}{Bandy}{2021}]%
        {bandy2021problematic}
\bibfield{author}{\bibinfo{person}{Jack Bandy}.}
  \bibinfo{year}{2021}\natexlab{}.
\newblock \showarticletitle{Problematic Machine Behavior: A Systematic
  Literature Review of Algorithm Audits}.
\newblock \bibinfo{journal}{\emph{Proceedings of the ACM on Human-Computer
  Interaction}} \bibinfo{volume}{5}, \bibinfo{number}{CSCW1}
  (\bibinfo{year}{2021}), \bibinfo{pages}{1--34}.
\newblock


\bibitem[\protect\citeauthoryear{Barocas, Guo, Kamar, Krones, Morris, Vaughan,
  Wadsworth, and Wallach}{Barocas et~al\mbox{.}}{2021}]%
        {barocas2021designing}
\bibfield{author}{\bibinfo{person}{Solon Barocas}, \bibinfo{person}{Anhong
  Guo}, \bibinfo{person}{Ece Kamar}, \bibinfo{person}{Jacquelyn Krones},
  \bibinfo{person}{Meredith~Ringel Morris}, \bibinfo{person}{Jennifer~Wortman
  Vaughan}, \bibinfo{person}{Duncan Wadsworth}, {and} \bibinfo{person}{Hanna
  Wallach}.} \bibinfo{year}{2021}\natexlab{}.
\newblock \bibinfo{title}{Designing Disaggregated Evaluations of AI Systems:
  Choices, Considerations, and Tradeoffs}.
\newblock
\newblock
\showeprint{arXiv:2103.06076}


\bibitem[\protect\citeauthoryear{Barocas, Hardt, and Narayanan}{Barocas
  et~al\mbox{.}}{2017}]%
        {barocas2017fairness}
\bibfield{author}{\bibinfo{person}{Solon Barocas}, \bibinfo{person}{Moritz
  Hardt}, {and} \bibinfo{person}{Arvind Narayanan}.}
  \bibinfo{year}{2017}\natexlab{}.
\newblock \showarticletitle{Fairness in Machine Learning}.
\newblock \bibinfo{journal}{\emph{Nips Tutorial}}  \bibinfo{volume}{1}
  (\bibinfo{year}{2017}), \bibinfo{pages}{2017}.
\newblock


\bibitem[\protect\citeauthoryear{Bendick~Jr and Nunes}{Bendick~Jr and
  Nunes}{2012}]%
        {bendick2012developing}
\bibfield{author}{\bibinfo{person}{Marc Bendick~Jr} {and}
  \bibinfo{person}{Ana~P Nunes}.} \bibinfo{year}{2012}\natexlab{}.
\newblock \showarticletitle{Developing the Research Basis for Controlling Bias
  in Hiring}.
\newblock \bibinfo{journal}{\emph{Journal of Social Issues}}
  \bibinfo{volume}{68}, \bibinfo{number}{2} (\bibinfo{year}{2012}),
  \bibinfo{pages}{238--262}.
\newblock


\bibitem[\protect\citeauthoryear{Bertrand and Duflo}{Bertrand and
  Duflo}{2017}]%
        {bertrand2017field}
\bibfield{author}{\bibinfo{person}{Marianne Bertrand} {and}
  \bibinfo{person}{Esther Duflo}.} \bibinfo{year}{2017}\natexlab{}.
\newblock \showarticletitle{Field Experiments on Discrimination}.
\newblock \bibinfo{journal}{\emph{Handbook of Economic Field Experiments}}
  \bibinfo{volume}{1} (\bibinfo{year}{2017}), \bibinfo{pages}{309--393}.
\newblock


\bibitem[\protect\citeauthoryear{Bertrand and Mullainathan}{Bertrand and
  Mullainathan}{2004}]%
        {bertrand2004emily}
\bibfield{author}{\bibinfo{person}{Marianne Bertrand} {and}
  \bibinfo{person}{Sendhil Mullainathan}.} \bibinfo{year}{2004}\natexlab{}.
\newblock \showarticletitle{Are Emily and Greg more Employable than Lakisha and
  Jamal? A Field Experiment on Labor Market Discrimination}.
\newblock \bibinfo{journal}{\emph{American Economic Review}}
  \bibinfo{volume}{94}, \bibinfo{number}{4} (\bibinfo{year}{2004}),
  \bibinfo{pages}{991--1013}.
\newblock


\bibitem[\protect\citeauthoryear{Besbris, Faber, Rich, and Sharkey}{Besbris
  et~al\mbox{.}}{2018}]%
        {besbris2018geography}
\bibfield{author}{\bibinfo{person}{Max Besbris}, \bibinfo{person}{Jacob~William
  Faber}, \bibinfo{person}{Peter Rich}, {and} \bibinfo{person}{Patrick
  Sharkey}.} \bibinfo{year}{2018}\natexlab{}.
\newblock \showarticletitle{The Geography of Stigma: Experimental Methods to
  Identify the Penalty of Place}.
\newblock In \bibinfo{booktitle}{\emph{Audit Studies: Behind the Scenes with
  Theory, Method, and Nuance}}. \bibinfo{publisher}{Springer},
  \bibinfo{address}{Berlin, Germany}, \bibinfo{pages}{159--177}.
\newblock


\bibitem[\protect\citeauthoryear{Bogen and Rieke}{Bogen and Rieke}{2018}]%
        {bogen2018help}
\bibfield{author}{\bibinfo{person}{Miranda Bogen} {and} \bibinfo{person}{Aaron
  Rieke}.} \bibinfo{year}{2018}\natexlab{}.
\newblock \bibinfo{title}{Help Wanted: An Examination of Hiring Algorithms,
  Equity, and Bias}.
\newblock
  \bibinfo{howpublished}{\url{https://www.upturn.org/reports/2018/hiring-algorithms/}}.
\newblock


\bibitem[\protect\citeauthoryear{Boyd, Levy, and Marwick}{Boyd
  et~al\mbox{.}}{2014}]%
        {boyd2014networked}
\bibfield{author}{\bibinfo{person}{Danah Boyd}, \bibinfo{person}{Karen Levy},
  {and} \bibinfo{person}{Alice Marwick}.} \bibinfo{year}{2014}\natexlab{}.
\newblock \bibinfo{title}{The Networked Nature of Algorithmic Discrimination}.
\newblock , \bibinfo{numpages}{53--57}~pages.
\newblock


\bibitem[\protect\citeauthoryear{Brown, Davidovic, and Hasan}{Brown
  et~al\mbox{.}}{2021}]%
        {brown2021algorithm}
\bibfield{author}{\bibinfo{person}{Shea Brown}, \bibinfo{person}{Jovana
  Davidovic}, {and} \bibinfo{person}{Ali Hasan}.}
  \bibinfo{year}{2021}\natexlab{}.
\newblock \showarticletitle{The Algorithm Audit: Scoring the Algorithms that
  Score us}.
\newblock \bibinfo{journal}{\emph{Big Data \& Society}} \bibinfo{volume}{8},
  \bibinfo{number}{1} (\bibinfo{year}{2021}), \bibinfo{pages}{1--12}.
\newblock


\bibitem[\protect\citeauthoryear{Buolamwini}{Buolamwini}{2018}]%
        {NYT}
\bibfield{author}{\bibinfo{person}{Joy Buolamwini}.}
  \bibinfo{year}{2018}\natexlab{}.
\newblock \bibinfo{title}{{When the Robot Doesn’t See Dark Skin}}.
\newblock
  \bibinfo{howpublished}{\url{https://www.nytimes.com/2018/06/21/opinion/facial-analysis-technology-bias.html}}.
\newblock


\bibitem[\protect\citeauthoryear{Buolamwini and Gebru}{Buolamwini and
  Gebru}{2018}]%
        {buolamwini2018gender}
\bibfield{author}{\bibinfo{person}{Joy Buolamwini} {and}
  \bibinfo{person}{Timnit Gebru}.} \bibinfo{year}{2018}\natexlab{}.
\newblock \showarticletitle{Gender shades: Intersectional Accuracy Disparities
  in Commercial Gender Classification}. In \bibinfo{booktitle}{\emph{Conference
  on Fairness, Accountability and Transparency}}. \bibinfo{publisher}{ACM},
  \bibinfo{address}{New York, NY}, \bibinfo{pages}{77--91}.
\newblock


\bibitem[\protect\citeauthoryear{Cherry and Bendick}{Cherry and
  Bendick}{2018}]%
        {cherry2018making}
\bibfield{author}{\bibinfo{person}{Frances Cherry} {and} \bibinfo{person}{Marc
  Bendick}.} \bibinfo{year}{2018}\natexlab{}.
\newblock \showarticletitle{Making it count: Discrimination Auditing and the
  Activist Scholar Tradition}.
\newblock In \bibinfo{booktitle}{\emph{Audit Studies: Behind the Scenes with
  Theory, Method, and Nuance}}. \bibinfo{publisher}{Springer},
  \bibinfo{address}{Berlin, Germany}, \bibinfo{pages}{45--62}.
\newblock


\bibitem[\protect\citeauthoryear{Chowdhury and Williams}{Chowdhury and
  Williams}{2021}]%
        {TwitterBugBounty}
\bibfield{author}{\bibinfo{person}{Rumman Chowdhury} {and}
  \bibinfo{person}{Jutta Williams}.} \bibinfo{year}{2021}\natexlab{}.
\newblock \bibinfo{title}{Introducing Twitter’s First Algorithmic Bias Bounty
  Challenge}.
\newblock
  \bibinfo{howpublished}{\url{https://blog.twitter.com/engineering/en_us/topics/insights/2021/algorithmic-bias-bounty-challenge}}.
\newblock


\bibitem[\protect\citeauthoryear{Crabtree and Chykina}{Crabtree and
  Chykina}{2018}]%
        {crabtree2018last}
\bibfield{author}{\bibinfo{person}{Charles Crabtree} {and}
  \bibinfo{person}{Volha Chykina}.} \bibinfo{year}{2018}\natexlab{}.
\newblock \showarticletitle{Last Name Selection in Audit Studies}.
\newblock \bibinfo{journal}{\emph{Sociological Science}}  \bibinfo{volume}{5}
  (\bibinfo{year}{2018}), \bibinfo{pages}{21--28}.
\newblock


\bibitem[\protect\citeauthoryear{Dastin}{Dastin}{2018}]%
        {Dastin_2018}
\bibfield{author}{\bibinfo{person}{Jeffrey Dastin}.}
  \bibinfo{year}{2018}\natexlab{}.
\newblock \bibinfo{title}{Amazon Scraps Secret AI Recruiting Tool that Showed
  Bias against Women}.
\newblock
  \bibinfo{howpublished}{\url{https://www.reuters.com/article/us-amazon-com-jobs-automation-insight-idUSKCN1MK08G}}.
\newblock


\bibitem[\protect\citeauthoryear{Datta, Tschantz, and Datta}{Datta
  et~al\mbox{.}}{2015}]%
        {datta2015automated}
\bibfield{author}{\bibinfo{person}{Amit Datta}, \bibinfo{person}{Michael~Carl
  Tschantz}, {and} \bibinfo{person}{Anupam Datta}.}
  \bibinfo{year}{2015}\natexlab{}.
\newblock \showarticletitle{Automated Experiments on Ad Privacy Settings: A
  Tale of Opacity, Choice, and Discrimination}.
\newblock \bibinfo{journal}{\emph{Proceedings on Privacy Enhancing
  Technologies}} \bibinfo{volume}{2015}, \bibinfo{number}{1}
  (\bibinfo{year}{2015}), \bibinfo{pages}{92--112}.
\newblock


\bibitem[\protect\citeauthoryear{Diakopoulos}{Diakopoulos}{2015}]%
        {diakopoulos2015algorithmic}
\bibfield{author}{\bibinfo{person}{Nicholas Diakopoulos}.}
  \bibinfo{year}{2015}\natexlab{}.
\newblock \showarticletitle{Algorithmic Accountability: Journalistic
  Investigation of Computational Power Structures}.
\newblock \bibinfo{journal}{\emph{Digital journalism}} \bibinfo{volume}{3},
  \bibinfo{number}{3} (\bibinfo{year}{2015}), \bibinfo{pages}{398--415}.
\newblock


\bibitem[\protect\citeauthoryear{D'Ignazio and Klein}{D'Ignazio and
  Klein}{2020}]%
        {d2020data}
\bibfield{author}{\bibinfo{person}{Catherine D'Ignazio} {and}
  \bibinfo{person}{Lauren~F Klein}.} \bibinfo{year}{2020}\natexlab{}.
\newblock \bibinfo{booktitle}{\emph{Data Feminism}}.
\newblock \bibinfo{publisher}{MIT Press}, \bibinfo{address}{Cambridge, MA}.
\newblock


\bibitem[\protect\citeauthoryear{Ehrenkranz}{Ehrenkranz}{2019}]%
        {gizmodoFaceRec}
\bibfield{author}{\bibinfo{person}{Melanie Ehrenkranz}.}
  \bibinfo{year}{2019}\natexlab{}.
\newblock \bibinfo{title}{{Amazon's Face Recognition Tech Once Again Pegs
  Politicians as Criminals}}.
\newblock
  \bibinfo{howpublished}{\url{https://gizmodo.com/amazons-face-recognition-tech-once-again-pegs-politicia-1837215790}}.
\newblock


\bibitem[\protect\citeauthoryear{Eisenhardt and Graebner}{Eisenhardt and
  Graebner}{2007}]%
        {eisenhardt2007theory}
\bibfield{author}{\bibinfo{person}{Kathleen~M Eisenhardt} {and}
  \bibinfo{person}{Melissa~E Graebner}.} \bibinfo{year}{2007}\natexlab{}.
\newblock \showarticletitle{Theory Building from Cases: Opportunities and
  Challenges}.
\newblock \bibinfo{journal}{\emph{Academy of Management Journal}}
  \bibinfo{volume}{50}, \bibinfo{number}{1} (\bibinfo{year}{2007}),
  \bibinfo{pages}{25--32}.
\newblock


\bibitem[\protect\citeauthoryear{Elazari Bar~On}{Elazari Bar~On}{2018}]%
        {vice}
\bibfield{author}{\bibinfo{person}{Amit Elazari Bar~On}.}
  \bibinfo{year}{2018}\natexlab{}.
\newblock \bibinfo{title}{{We Need Bug Bounties for Bad Algorithms}}.
\newblock
  \bibinfo{howpublished}{\url{https://www.vice.com/en/article/8xkyj3/we-need-bug-bounties-for-bad-algorithms}}.
\newblock


\bibitem[\protect\citeauthoryear{Eslami, Vaccaro, Karahalios, and
  Hamilton}{Eslami et~al\mbox{.}}{2017}]%
        {eslami2017careful}
\bibfield{author}{\bibinfo{person}{Motahhare Eslami}, \bibinfo{person}{Kristen
  Vaccaro}, \bibinfo{person}{Karrie Karahalios}, {and} \bibinfo{person}{Kevin
  Hamilton}.} \bibinfo{year}{2017}\natexlab{}.
\newblock \showarticletitle{“Be Careful; Things can be Worse than they
  Appear”: Understanding Biased Algorithms and Users’ Behavior around Them
  in Rating Platforms}. In \bibinfo{booktitle}{\emph{Proceedings of the
  International AAAI Conference on Web and Social Media}},
  Vol.~\bibinfo{volume}{11}. \bibinfo{publisher}{AAAI Press},
  \bibinfo{address}{Palo Alto, California}, \bibinfo{pages}{62--71}.
\newblock


\bibitem[\protect\citeauthoryear{Gaddis}{Gaddis}{2017}]%
        {gaddis2017black}
\bibfield{author}{\bibinfo{person}{S~Michael Gaddis}.}
  \bibinfo{year}{2017}\natexlab{}.
\newblock \showarticletitle{How Black are Lakisha and Jamal? Racial Perceptions
  from Names used in Correspondence Audit Studies}.
\newblock \bibinfo{journal}{\emph{Sociological Science}}  \bibinfo{volume}{4}
  (\bibinfo{year}{2017}), \bibinfo{pages}{469--489}.
\newblock


\bibitem[\protect\citeauthoryear{Gaddis}{Gaddis}{2018}]%
        {gaddis2018introduction}
\bibfield{author}{\bibinfo{person}{S~Michael Gaddis}.}
  \bibinfo{year}{2018}\natexlab{}.
\newblock \bibinfo{booktitle}{\emph{An Introduction to Audit Studies in the
  Social Sciences}}.
\newblock \bibinfo{publisher}{Springer}, \bibinfo{address}{Berlin, Germany}.
  3--44 pages.
\newblock


\bibitem[\protect\citeauthoryear{Garvie}{Garvie}{2019}]%
        {garbageInGarbageOut}
\bibfield{author}{\bibinfo{person}{Clare Garvie}.}
  \bibinfo{year}{2019}\natexlab{}.
\newblock \bibinfo{title}{{Garbage In, Garbage Out}}.
\newblock \bibinfo{howpublished}{\url{https://www.flawedfacedata.com/}}.
\newblock


\bibitem[\protect\citeauthoryear{Gell-Redman, Visalvanich, Crabtree, and
  Fariss}{Gell-Redman et~al\mbox{.}}{2018}]%
        {gell2018s}
\bibfield{author}{\bibinfo{person}{Micah Gell-Redman}, \bibinfo{person}{Neil
  Visalvanich}, \bibinfo{person}{Charles Crabtree}, {and}
  \bibinfo{person}{Christopher~J Fariss}.} \bibinfo{year}{2018}\natexlab{}.
\newblock \showarticletitle{It’s all About Race: How State Legislators
  Respond to Immigrant Constituents}.
\newblock \bibinfo{journal}{\emph{Political Research Quarterly}}
  \bibinfo{volume}{71}, \bibinfo{number}{3} (\bibinfo{year}{2018}),
  \bibinfo{pages}{517--531}.
\newblock


\bibitem[\protect\citeauthoryear{Granovetter}{Granovetter}{1995}]%
        {granovetter1995getting}
\bibfield{author}{\bibinfo{person}{Mark Granovetter}.}
  \bibinfo{year}{1995}\natexlab{}.
\newblock \bibinfo{booktitle}{\emph{Getting a Job: A Study of Contacts and
  Careers}}.
\newblock \bibinfo{publisher}{University of Chicago Press},
  \bibinfo{address}{Chicago, IL}.
\newblock


\bibitem[\protect\citeauthoryear{Gunawardana and Shani}{Gunawardana and
  Shani}{2009}]%
        {gunawardana2009survey}
\bibfield{author}{\bibinfo{person}{Asela Gunawardana} {and}
  \bibinfo{person}{Guy Shani}.} \bibinfo{year}{2009}\natexlab{}.
\newblock \showarticletitle{A Survey of Accuracy Evaluation Metrics of
  Recommendation Tasks}.
\newblock \bibinfo{journal}{\emph{Journal of Machine Learning Research}}
  \bibinfo{volume}{10}, \bibinfo{number}{12} (\bibinfo{year}{2009}),
  \bibinfo{pages}{2936--2962}.
\newblock


\bibitem[\protect\citeauthoryear{Guryan and Charles}{Guryan and
  Charles}{2013}]%
        {guryan2013taste}
\bibfield{author}{\bibinfo{person}{Jonathan Guryan} {and}
  \bibinfo{person}{Kerwin~Kofi Charles}.} \bibinfo{year}{2013}\natexlab{}.
\newblock \showarticletitle{Taste-Based or Statistical Discrimination: the
  Economics of Discrimination Returns to its Roots}.
\newblock \bibinfo{journal}{\emph{The Economic Journal}} \bibinfo{volume}{123},
  \bibinfo{number}{572} (\bibinfo{year}{2013}), \bibinfo{pages}{F417--F432}.
\newblock


\bibitem[\protect\citeauthoryear{Hann{\'a}k, Wagner, Garcia, Mislove,
  Strohmaier, and Wilson}{Hann{\'a}k et~al\mbox{.}}{2017}]%
        {hannak2017bias}
\bibfield{author}{\bibinfo{person}{Anik{\'o} Hann{\'a}k},
  \bibinfo{person}{Claudia Wagner}, \bibinfo{person}{David Garcia},
  \bibinfo{person}{Alan Mislove}, \bibinfo{person}{Markus Strohmaier}, {and}
  \bibinfo{person}{Christo Wilson}.} \bibinfo{year}{2017}\natexlab{}.
\newblock \showarticletitle{Bias in Online Freelance Marketplaces: Evidence
  from Taskrabbit and Fiverr}. In \bibinfo{booktitle}{\emph{Proceedings of the
  2017 ACM Conference on Computer Supported Cooperative Work and Social
  Computing}}. \bibinfo{publisher}{ACM}, \bibinfo{address}{New York, NY},
  \bibinfo{pages}{1914--1933}.
\newblock


\bibitem[\protect\citeauthoryear{Hansson}{Hansson}{2019}]%
        {dhh}
\bibfield{author}{\bibinfo{person}{David~Heinemeier Hansson}.}
  \bibinfo{year}{2019}\natexlab{}.
\newblock \bibinfo{title}{{The AppleCard is such a fucking sexist program. My
  wife and I filed joint tax returns, live in a community-property state, and
  have been married for a long time. Yet Apple’s black box algorithm thinks I
  deserve 20x the credit limit she does. No appeals work.}}
\newblock
  \bibinfo{howpublished}{\url{https://twitter.com/dhh/status/1192540900393705474}}.
\newblock


\bibitem[\protect\citeauthoryear{Hoffmann}{Hoffmann}{2019}]%
        {hoffmann2019fairness}
\bibfield{author}{\bibinfo{person}{Anna~Lauren Hoffmann}.}
  \bibinfo{year}{2019}\natexlab{}.
\newblock \showarticletitle{Where Fairness Fails: Data, Algorithms, and the
  Limits of Antidiscrimination Discourse}.
\newblock \bibinfo{journal}{\emph{Information, Communication \& Society}}
  \bibinfo{volume}{22}, \bibinfo{number}{7} (\bibinfo{year}{2019}),
  \bibinfo{pages}{900--915}.
\newblock


\bibitem[\protect\citeauthoryear{Hu and Kohler-Hausmann}{Hu and
  Kohler-Hausmann}{2020}]%
        {hu2020s}
\bibfield{author}{\bibinfo{person}{Lily Hu} {and} \bibinfo{person}{Issa
  Kohler-Hausmann}.} \bibinfo{year}{2020}\natexlab{}.
\newblock \bibinfo{title}{What's Sex Got To Do With Fair Machine Learning}.
\newblock
\newblock
\showeprint{arXiv:2006.01770}


\bibitem[\protect\citeauthoryear{Imana, Korolova, and Heidemann}{Imana
  et~al\mbox{.}}{2021}]%
        {imana2021auditing}
\bibfield{author}{\bibinfo{person}{Basileal Imana}, \bibinfo{person}{Aleksandra
  Korolova}, {and} \bibinfo{person}{John Heidemann}.}
  \bibinfo{year}{2021}\natexlab{}.
\newblock \showarticletitle{Auditing for Discrimination in Algorithms
  Delivering Job Ads}. In \bibinfo{booktitle}{\emph{Proceedings of the Web
  Conference 2021}}. \bibinfo{publisher}{ACM}, \bibinfo{address}{New York, NY},
  \bibinfo{pages}{3767--3778}.
\newblock


\bibitem[\protect\citeauthoryear{Introna and Nissenbaum}{Introna and
  Nissenbaum}{2010}]%
        {introna2010facial}
\bibfield{author}{\bibinfo{person}{Lucas Introna} {and} \bibinfo{person}{Helen
  Nissenbaum}.} \bibinfo{year}{2010}\natexlab{}.
\newblock \bibinfo{title}{Facial Recognition Technology a Survey of Policy and
  Implementation Issues}.
\newblock , \bibinfo{numpages}{8--55}~pages.
\newblock


\bibitem[\protect\citeauthoryear{{Jason Slotkin}}{{Jason Slotkin}}{2020}]%
        {NPR}
\bibfield{author}{\bibinfo{person}{{Jason Slotkin}}.}
  \bibinfo{year}{2020}\natexlab{}.
\newblock \bibinfo{title}{{Twitter Announces Changes To Image Cropping Amid
  Bias Concern}}.
\newblock
  \bibinfo{howpublished}{\url{https://www.npr.org/sections/live-updates-protests-for-racial-justice/2020/10/02/919638417/twitter-announces-changes-to-image-cropping-amid-bias-concern}}.
\newblock


\bibitem[\protect\citeauthoryear{Kroll, Huey, Barocas, Felten, Reidenberg,
  Robinson, and Yu}{Kroll et~al\mbox{.}}{2017}]%
        {kroll2016accountable}
\bibfield{author}{\bibinfo{person}{Joshua Kroll}, \bibinfo{person}{Joanna
  Huey}, \bibinfo{person}{Solon Barocas}, \bibinfo{person}{Edward Felten},
  \bibinfo{person}{Joel Reidenberg}, \bibinfo{person}{David Robinson}, {and}
  \bibinfo{person}{Harlan Yu}.} \bibinfo{year}{2017}\natexlab{}.
\newblock \showarticletitle{Accountable Algorithms}.
\newblock \bibinfo{journal}{\emph{University of Pennsylvania Law Review}}
  \bibinfo{volume}{165}, \bibinfo{number}{3} (\bibinfo{year}{2017}),
  \bibinfo{pages}{633--705}.
\newblock


\bibitem[\protect\citeauthoryear{Mahieu and Ausloos}{Mahieu and
  Ausloos}{2020}]%
        {mahieu2020recognising}
\bibfield{author}{\bibinfo{person}{Ren{\'e} Mahieu} {and} \bibinfo{person}{Jef
  Ausloos}.} \bibinfo{year}{2020}\natexlab{}.
\newblock \bibinfo{title}{Recognising and Enabling the Collective Dimension of
  the GDPR and the Right of Access}.
\newblock
\newblock


\bibitem[\protect\citeauthoryear{{Manish Raghavan and Solon Barocas}}{{Manish
  Raghavan and Solon Barocas}}{2019}]%
        {raghavan2019challenges}
\bibfield{author}{\bibinfo{person}{{Manish Raghavan and Solon Barocas}}.}
  \bibinfo{year}{2019}\natexlab{}.
\newblock \bibinfo{title}{{Challenges for Mitigating Bias in Algorithmic
  Hiring}}.
\newblock
  \bibinfo{howpublished}{\url{https://www.brookings.edu/research/challenges-for-mitigating-bias-in-algorithmic-hiring/}}.
\newblock


\bibitem[\protect\citeauthoryear{Markup}{Markup}{2020}]%
        {markup2020}
\bibfield{author}{\bibinfo{person}{The Markup}.}
  \bibinfo{year}{2020}\natexlab{}.
\newblock \bibinfo{title}{The Citizen Browser Project: Auditing the Algorithms
  of Disinformation}.
\newblock \bibinfo{howpublished}{\url{https://themarkup.org/citizen-browser}}.
\newblock


\bibitem[\protect\citeauthoryear{Martinez and Kirchner}{Martinez and
  Kirchner}{2021}]%
        {markup2021mortgage}
\bibfield{author}{\bibinfo{person}{Emmanuel Martinez} {and}
  \bibinfo{person}{Lauren Kirchner}.} \bibinfo{year}{2021}\natexlab{}.
\newblock \bibinfo{title}{The Secret Bias Hidden in Mortgage-Approval
  Algorithms}.
\newblock
  \bibinfo{howpublished}{\url{https://themarkup.org/denied/2021/08/25/the-secret-bias-hidden-in-mortgage-approval-algorithms}}.
\newblock


\bibitem[\protect\citeauthoryear{McIntyre}{McIntyre}{2007}]%
        {mcintyre2007participatory}
\bibfield{author}{\bibinfo{person}{Alice McIntyre}.}
  \bibinfo{year}{2007}\natexlab{}.
\newblock \bibinfo{booktitle}{\emph{Participatory Action Research}}.
\newblock \bibinfo{publisher}{Sage Publications}, \bibinfo{address}{New York,
  NY}.
\newblock


\bibitem[\protect\citeauthoryear{McPherson, Smith-Lovin, and Cook}{McPherson
  et~al\mbox{.}}{2001}]%
        {mcpherson2001birds}
\bibfield{author}{\bibinfo{person}{Miller McPherson}, \bibinfo{person}{Lynn
  Smith-Lovin}, {and} \bibinfo{person}{James~M Cook}.}
  \bibinfo{year}{2001}\natexlab{}.
\newblock \showarticletitle{Birds of a Feather: Homophily in Social Networks}.
\newblock \bibinfo{journal}{\emph{Annual Review of Sociology}}
  \bibinfo{volume}{27}, \bibinfo{number}{1} (\bibinfo{year}{2001}),
  \bibinfo{pages}{415--444}.
\newblock


\bibitem[\protect\citeauthoryear{Mishel}{Mishel}{2016}]%
        {mishel2016discrimination}
\bibfield{author}{\bibinfo{person}{Emma Mishel}.}
  \bibinfo{year}{2016}\natexlab{}.
\newblock \showarticletitle{Discrimination Against Queer Women in the US
  Workforce: A R{\'e}sum{\'e} Audit Study}.
\newblock \bibinfo{journal}{\emph{Socius}}  \bibinfo{volume}{2}
  (\bibinfo{year}{2016}), \bibinfo{pages}{1--13}.
\newblock


\bibitem[\protect\citeauthoryear{Mittelstadt}{Mittelstadt}{2016}]%
        {mittelstadt2016automation}
\bibfield{author}{\bibinfo{person}{Brent Mittelstadt}.}
  \bibinfo{year}{2016}\natexlab{}.
\newblock \showarticletitle{Auditing for Transparency in Content
  Personalization Systems}.
\newblock \bibinfo{journal}{\emph{International Journal of Communication}}
  \bibinfo{volume}{10} (\bibinfo{year}{2016}), \bibinfo{pages}{4991--5002}.
\newblock


\bibitem[\protect\citeauthoryear{Nasiripour and Farrell}{Nasiripour and
  Farrell}{2021}]%
        {Nasiripour_Farrell_2021}
\bibfield{author}{\bibinfo{person}{Shahien Nasiripour} {and}
  \bibinfo{person}{Greg Farrell}.} \bibinfo{year}{2021}\natexlab{}.
\newblock \bibinfo{title}{Goldman Cleared of Bias in New York Review of Apple
  Card}.
\newblock
  \bibinfo{howpublished}{\url{https://www.bloomberg.com/news/articles/2021-03-23/goldman-didn-t-discriminate-with-apple-card-n-y-regulator-says}}.
\newblock


\bibitem[\protect\citeauthoryear{Ng}{Ng}{2021}]%
        {Ng_2021}
\bibfield{author}{\bibinfo{person}{Alfred Ng}.}
  \bibinfo{year}{2021}\natexlab{}.
\newblock \bibinfo{title}{Can Auditing Eliminate Bias from Algorithms?}
\newblock
\newblock
\urldef\tempurl%
\url{https://themarkup.org/ask-the-markup/2021/02/23/can-auditing-eliminate-bias-from-algorithms}
\showURL{%
\tempurl}


\bibitem[\protect\citeauthoryear{Pager}{Pager}{2007}]%
        {pager2007use}
\bibfield{author}{\bibinfo{person}{Devah Pager}.}
  \bibinfo{year}{2007}\natexlab{}.
\newblock \showarticletitle{The Use of Field Experiments for Studies of
  Employment discrimination: Contributions, Critiques, and Directions for the
  Future}.
\newblock \bibinfo{journal}{\emph{The Annals of the American Academy of
  Political and Social Science}} \bibinfo{volume}{609}, \bibinfo{number}{1}
  (\bibinfo{year}{2007}), \bibinfo{pages}{104--133}.
\newblock


\bibitem[\protect\citeauthoryear{{Parag Agrawal and Dantley Davis}}{{Parag
  Agrawal and Dantley Davis}}{2020}]%
        {twitterTransparency}
\bibfield{author}{\bibinfo{person}{{Parag Agrawal and Dantley Davis}}.}
  \bibinfo{year}{2020}\natexlab{}.
\newblock \bibinfo{title}{{Transparency around Image Cropping and Changes to
  Come}}.
\newblock
  \bibinfo{howpublished}{\url{https://blog.twitter.com/official/en_us/topics/product/2020/transparency-image-cropping.html}}.
\newblock


\bibitem[\protect\citeauthoryear{Pedulla}{Pedulla}{2016}]%
        {pedulla2016penalized}
\bibfield{author}{\bibinfo{person}{David~S Pedulla}.}
  \bibinfo{year}{2016}\natexlab{}.
\newblock \showarticletitle{Penalized or Protected? Gender and the Consequences
  of Nonstandard and Mismatched Employment Histories}.
\newblock \bibinfo{journal}{\emph{American Sociological Review}}
  \bibinfo{volume}{81}, \bibinfo{number}{2} (\bibinfo{year}{2016}),
  \bibinfo{pages}{262--289}.
\newblock


\bibitem[\protect\citeauthoryear{Pedulla}{Pedulla}{2018}]%
        {pedulla2018emerging}
\bibfield{author}{\bibinfo{person}{David~S Pedulla}.}
  \bibinfo{year}{2018}\natexlab{}.
\newblock \showarticletitle{Emerging Frontiers in Audit Study Research:
  Mechanisms, Variation, and Representativeness}.
\newblock In \bibinfo{booktitle}{\emph{Audit Studies: Behind the Scenes with
  Theory, Method, and Nuance}}. \bibinfo{publisher}{Springer},
  \bibinfo{address}{New York, NY}, \bibinfo{pages}{179--195}.
\newblock


\bibitem[\protect\citeauthoryear{Quillian, Lee, and Oliver}{Quillian
  et~al\mbox{.}}{2020}]%
        {quillian2020evidence}
\bibfield{author}{\bibinfo{person}{Lincoln Quillian}, \bibinfo{person}{John~J
  Lee}, {and} \bibinfo{person}{Mariana Oliver}.}
  \bibinfo{year}{2020}\natexlab{}.
\newblock \showarticletitle{Evidence From Field Experiments in Hiring Shows
  Substantial Additional Racial Discrimination after the Callback}.
\newblock \bibinfo{journal}{\emph{Social Forces}} \bibinfo{volume}{99},
  \bibinfo{number}{2} (\bibinfo{year}{2020}), \bibinfo{pages}{732--759}.
\newblock


\bibitem[\protect\citeauthoryear{Raghavan, Barocas, Kleinberg, and
  Levy}{Raghavan et~al\mbox{.}}{2020}]%
        {raghavan2020mitigating}
\bibfield{author}{\bibinfo{person}{Manish Raghavan}, \bibinfo{person}{Solon
  Barocas}, \bibinfo{person}{Jon Kleinberg}, {and} \bibinfo{person}{Karen
  Levy}.} \bibinfo{year}{2020}\natexlab{}.
\newblock \showarticletitle{Mitigating Bias in Algorithmic Hiring: Evaluating
  Claims and Practices}. In \bibinfo{booktitle}{\emph{Proceedings of the 2020
  Conference on Fairness, Accountability, and Transparency}}.
  \bibinfo{publisher}{ACM}, \bibinfo{address}{New York, NY},
  \bibinfo{pages}{469--481}.
\newblock


\bibitem[\protect\citeauthoryear{Raji, Gebru, Mitchell, Buolamwini, Lee, and
  Denton}{Raji et~al\mbox{.}}{2020a}]%
        {raji2020saving}
\bibfield{author}{\bibinfo{person}{Inioluwa~Deborah Raji},
  \bibinfo{person}{Timnit Gebru}, \bibinfo{person}{Margaret Mitchell},
  \bibinfo{person}{Joy Buolamwini}, \bibinfo{person}{Joonseok Lee}, {and}
  \bibinfo{person}{Emily Denton}.} \bibinfo{year}{2020}\natexlab{a}.
\newblock \showarticletitle{Saving Face: Investigating the Ethical Concerns of
  Facial Recognition Auditing}. In \bibinfo{booktitle}{\emph{Proceedings of the
  AAAI/ACM Conference on AI, Ethics, and Society}}. \bibinfo{publisher}{ACM},
  \bibinfo{address}{New York, NY}, \bibinfo{pages}{145--151}.
\newblock


\bibitem[\protect\citeauthoryear{Raji, Smart, White, Mitchell, Gebru,
  Hutchinson, Smith-Loud, Theron, and Barnes}{Raji et~al\mbox{.}}{2020b}]%
        {raji2020closing}
\bibfield{author}{\bibinfo{person}{Inioluwa~Deborah Raji},
  \bibinfo{person}{Andrew Smart}, \bibinfo{person}{Rebecca~N White},
  \bibinfo{person}{Margaret Mitchell}, \bibinfo{person}{Timnit Gebru},
  \bibinfo{person}{Ben Hutchinson}, \bibinfo{person}{Jamila Smith-Loud},
  \bibinfo{person}{Daniel Theron}, {and} \bibinfo{person}{Parker Barnes}.}
  \bibinfo{year}{2020}\natexlab{b}.
\newblock \showarticletitle{Closing the AI Accountability Gap: Defining an
  End-to-End Framework for Internal Algorithmic Auditing}. In
  \bibinfo{booktitle}{\emph{Proceedings of the 2020 Conference on Fairness,
  Accountability, and Transparency}}. \bibinfo{publisher}{ACM},
  \bibinfo{address}{New York, NY}, \bibinfo{pages}{33--44}.
\newblock


\bibitem[\protect\citeauthoryear{Reichborn-Kjennerud and
  Vabo}{Reichborn-Kjennerud and Vabo}{2017}]%
        {reichborn2017performance}
\bibfield{author}{\bibinfo{person}{Kristin Reichborn-Kjennerud} {and}
  \bibinfo{person}{Signy~Irene Vabo}.} \bibinfo{year}{2017}\natexlab{}.
\newblock \showarticletitle{Performance Audit as a Contributor to Change and
  Improvement in Public Administration}.
\newblock \bibinfo{journal}{\emph{Evaluation}} \bibinfo{volume}{23},
  \bibinfo{number}{1} (\bibinfo{year}{2017}), \bibinfo{pages}{6--23}.
\newblock


\bibitem[\protect\citeauthoryear{Rieke, Janardan, Hsu, and Duarte}{Rieke
  et~al\mbox{.}}{2021}]%
        {riekeupturn}
\bibfield{author}{\bibinfo{person}{Aaron Rieke}, \bibinfo{person}{Urmila
  Janardan}, \bibinfo{person}{Mingwei Hsu}, {and} \bibinfo{person}{Natasha
  Duarte}.} \bibinfo{year}{2021}\natexlab{}.
\newblock \bibinfo{title}{{Essential Work: Analyzing the Hiring Technologies of
  Large Hourly Employers}}.
\newblock
  \bibinfo{howpublished}{\url{https://www.upturn.org/reports/2021/essential-work/}}.
\newblock


\bibitem[\protect\citeauthoryear{Rivera}{Rivera}{2016}]%
        {rivera2016pedigree}
\bibfield{author}{\bibinfo{person}{Lauren~A Rivera}.}
  \bibinfo{year}{2016}\natexlab{}.
\newblock \bibinfo{booktitle}{\emph{Pedigree: How Elite Students get Elite
  Jobs}}.
\newblock \bibinfo{publisher}{Princeton University Press},
  \bibinfo{address}{Princeton, NJ}.
\newblock


\bibitem[\protect\citeauthoryear{Robertson, Jiang, Joseph, Friedland, Lazer,
  and Wilson}{Robertson et~al\mbox{.}}{2018}]%
        {robertson2018auditing}
\bibfield{author}{\bibinfo{person}{Ronald~E Robertson}, \bibinfo{person}{Shan
  Jiang}, \bibinfo{person}{Kenneth Joseph}, \bibinfo{person}{Lisa Friedland},
  \bibinfo{person}{David Lazer}, {and} \bibinfo{person}{Christo Wilson}.}
  \bibinfo{year}{2018}\natexlab{}.
\newblock \showarticletitle{Auditing Partisan Audience Bias within Google
  Search}.
\newblock \bibinfo{journal}{\emph{Proceedings of the ACM on Human-Computer
  Interaction}} \bibinfo{volume}{2}, \bibinfo{number}{CSCW}
  (\bibinfo{year}{2018}), \bibinfo{pages}{1--22}.
\newblock


\bibitem[\protect\citeauthoryear{Robertson, Olteanu, Diaz, Shokouhi, and
  Bailey}{Robertson et~al\mbox{.}}{2021}]%
        {robertson2021can}
\bibfield{author}{\bibinfo{person}{Ronald~E Robertson},
  \bibinfo{person}{Alexandra Olteanu}, \bibinfo{person}{Fernando Diaz},
  \bibinfo{person}{Milad Shokouhi}, {and} \bibinfo{person}{Peter Bailey}.}
  \bibinfo{year}{2021}\natexlab{}.
\newblock \showarticletitle{“I Can’t Reply with That”: Characterizing
  Problematic Email Reply Suggestions}. In
  \bibinfo{booktitle}{\emph{Proceedings of the 2021 CHI Conference on Human
  Factors in Computing Systems}}. \bibinfo{publisher}{ACM},
  \bibinfo{address}{New York, NY}, \bibinfo{pages}{1--18}.
\newblock


\bibitem[\protect\citeauthoryear{Salganik}{Salganik}{2019}]%
        {salganik2019bit}
\bibfield{author}{\bibinfo{person}{Matthew~J Salganik}.}
  \bibinfo{year}{2019}\natexlab{}.
\newblock \bibinfo{booktitle}{\emph{Bit by Bit: Social Research in the Digital
  Age}}.
\newblock \bibinfo{publisher}{Princeton University Press},
  \bibinfo{address}{Princeton, NJ}.
\newblock


\bibitem[\protect\citeauthoryear{Sandvig, Hamilton, Karahalios, and
  Langbort}{Sandvig et~al\mbox{.}}{2014}]%
        {sandvig2014auditing}
\bibfield{author}{\bibinfo{person}{Christian Sandvig}, \bibinfo{person}{Kevin
  Hamilton}, \bibinfo{person}{Karrie Karahalios}, {and} \bibinfo{person}{Cedric
  Langbort}.} \bibinfo{year}{2014}\natexlab{}.
\newblock \showarticletitle{Auditing Algorithms: Research Methods for Detecting
  Discrimination on Internet Platforms}.
\newblock \bibinfo{journal}{\emph{Data and discrimination: Converting Critical
  Concerns into Productive Inquiry}}  \bibinfo{volume}{22}
  (\bibinfo{year}{2014}), \bibinfo{pages}{4349--4357}.
\newblock


\bibitem[\protect\citeauthoryear{Saxena, Badillo-Urquiola, Wisniewski, and
  Guha}{Saxena et~al\mbox{.}}{2021}]%
        {saxena2021framework}
\bibfield{author}{\bibinfo{person}{Devansh Saxena}, \bibinfo{person}{Karla
  Badillo-Urquiola}, \bibinfo{person}{Pamela Wisniewski}, {and}
  \bibinfo{person}{Shion Guha}.} \bibinfo{year}{2021}\natexlab{}.
\newblock \bibinfo{title}{A Framework of High-Stakes Algorithmic
  Decision-Making for the Public Sector Developed through a Case Study of
  Child-Welfare}.
\newblock
\newblock
\showeprint{arXiv:2107.03487}


\bibitem[\protect\citeauthoryear{Selbst, Boyd, Friedler, Venkatasubramanian,
  and Vertesi}{Selbst et~al\mbox{.}}{2019}]%
        {selbst2019fairness}
\bibfield{author}{\bibinfo{person}{Andrew~D Selbst}, \bibinfo{person}{Danah
  Boyd}, \bibinfo{person}{Sorelle~A Friedler}, \bibinfo{person}{Suresh
  Venkatasubramanian}, {and} \bibinfo{person}{Janet Vertesi}.}
  \bibinfo{year}{2019}\natexlab{}.
\newblock \showarticletitle{Fairness and Abstraction in Sociotechnical
  Systems}. In \bibinfo{booktitle}{\emph{Proceedings of the Conference on
  Fairness, Accountability, and Transparency}}. \bibinfo{publisher}{ACM},
  \bibinfo{address}{New York, NY}, \bibinfo{pages}{59--68}.
\newblock


\bibitem[\protect\citeauthoryear{Sen and Wasow}{Sen and Wasow}{2016}]%
        {sen2016race}
\bibfield{author}{\bibinfo{person}{Maya Sen} {and} \bibinfo{person}{Omar
  Wasow}.} \bibinfo{year}{2016}\natexlab{}.
\newblock \showarticletitle{Race as a bundle of sticks: Designs that estimate
  effects of seemingly immutable characteristics}.
\newblock \bibinfo{journal}{\emph{Annual Review of Political Science}}
  \bibinfo{volume}{19} (\bibinfo{year}{2016}), \bibinfo{pages}{499--522}.
\newblock


\bibitem[\protect\citeauthoryear{Sloane}{Sloane}{2021}]%
        {Sloane_2021}
\bibfield{author}{\bibinfo{person}{Mona Sloane}.}
  \bibinfo{year}{2021}\natexlab{}.
\newblock \bibinfo{title}{The Algorithmic Auditing Trap}.
\newblock
  \bibinfo{howpublished}{\url{https://onezero.medium.com/the-algorithmic-auditing-trap-9a6f2d4d461d}}.
\newblock


\bibitem[\protect\citeauthoryear{Sloane, Moss, and Chowdhury}{Sloane
  et~al\mbox{.}}{2021}]%
        {sloane2021silicon}
\bibfield{author}{\bibinfo{person}{Mona Sloane}, \bibinfo{person}{Emanuel
  Moss}, {and} \bibinfo{person}{Rumman Chowdhury}.}
  \bibinfo{year}{2021}\natexlab{}.
\newblock \bibinfo{title}{A Silicon Valley Love Triangle: Hiring Algorithms,
  Pseudo-Science, and the Quest for Auditability}.
\newblock
\newblock
\showeprint{arXiv:2106.12403}


\bibitem[\protect\citeauthoryear{Small and Pager}{Small and Pager}{2020}]%
        {small2020sociological}
\bibfield{author}{\bibinfo{person}{Mario~L Small} {and} \bibinfo{person}{Devah
  Pager}.} \bibinfo{year}{2020}\natexlab{}.
\newblock \showarticletitle{Sociological Perspectives on Racial
  Discrimination}.
\newblock \bibinfo{journal}{\emph{Journal of Economic Perspectives}}
  \bibinfo{volume}{34}, \bibinfo{number}{2} (\bibinfo{year}{2020}),
  \bibinfo{pages}{49--67}.
\newblock


\bibitem[\protect\citeauthoryear{Snow}{Snow}{2018}]%
        {acluFaceRec}
\bibfield{author}{\bibinfo{person}{Jacob Snow}.}
  \bibinfo{year}{2018}\natexlab{}.
\newblock \bibinfo{title}{{Amazon’s Face Recognition Falsely Matched 28
  Members of Congress With Mugshots}}.
\newblock
  \bibinfo{howpublished}{\url{https://www.aclu.org/blog/privacy-technology/surveillance-technologies/amazons-face-recognition-falsely-matched-28}}.
\newblock


\bibitem[\protect\citeauthoryear{Stevenson}{Stevenson}{2018}]%
        {stevenson2018assessing}
\bibfield{author}{\bibinfo{person}{Megan Stevenson}.}
  \bibinfo{year}{2018}\natexlab{}.
\newblock \showarticletitle{Assessing Risk Assessment in Action}.
\newblock \bibinfo{journal}{\emph{Minnesota Law Review}}  \bibinfo{volume}{103}
  (\bibinfo{year}{2018}), \bibinfo{pages}{303}.
\newblock


\bibitem[\protect\citeauthoryear{Stevenson and Doleac}{Stevenson and
  Doleac}{2019}]%
        {IZAalgRiskAssessments}
\bibfield{author}{\bibinfo{person}{Megan~K. Stevenson} {and}
  \bibinfo{person}{Jennifer~L. Doleac}.} \bibinfo{year}{2019}\natexlab{}.
\newblock \bibinfo{booktitle}{\emph{{Algorithmic Risk Assessment in the Hands
  of Humans}}}.
\newblock \bibinfo{type}{{T}echnical {R}eport}. \bibinfo{institution}{IZA
  Institute of Labor Economics}.
\newblock


\bibitem[\protect\citeauthoryear{Strathern}{Strathern}{2000}]%
        {strathern2000audit}
\bibfield{author}{\bibinfo{person}{Marilyn Strathern}.}
  \bibinfo{year}{2000}\natexlab{}.
\newblock \bibinfo{booktitle}{\emph{Audit Cultures: Anthropological Studies in
  Accountability, Ethics, and the Academy}}.
\newblock \bibinfo{publisher}{Psychology Press}, \bibinfo{address}{Hove, East
  Sussex, UK}.
\newblock


\bibitem[\protect\citeauthoryear{Sullivan, Aycrigg, Barry, Bonney, Bruns,
  Cooper, Damoulas, Dhondt, Dietterich, Farnsworth, et~al\mbox{.}}{Sullivan
  et~al\mbox{.}}{2014}]%
        {sullivan2014ebird}
\bibfield{author}{\bibinfo{person}{Brian~L Sullivan},
  \bibinfo{person}{Jocelyn~L Aycrigg}, \bibinfo{person}{Jessie~H Barry},
  \bibinfo{person}{Rick~E Bonney}, \bibinfo{person}{Nicholas Bruns},
  \bibinfo{person}{Caren~B Cooper}, \bibinfo{person}{Theo Damoulas},
  \bibinfo{person}{Andr{\'e}~A Dhondt}, \bibinfo{person}{Tom Dietterich},
  \bibinfo{person}{Andrew Farnsworth}, {et~al\mbox{.}}}
  \bibinfo{year}{2014}\natexlab{}.
\newblock \showarticletitle{The eBird Enterprise: an Integrated Approach to
  Development and Application of Citizen Science}.
\newblock \bibinfo{journal}{\emph{Biological Conservation}}
  \bibinfo{volume}{169} (\bibinfo{year}{2014}), \bibinfo{pages}{31--40}.
\newblock


\bibitem[\protect\citeauthoryear{Sweeney}{Sweeney}{2013}]%
        {sweeney2013discrimination}
\bibfield{author}{\bibinfo{person}{Latanya Sweeney}.}
  \bibinfo{year}{2013}\natexlab{}.
\newblock \showarticletitle{Discrimination in Online Ad Delivery}.
\newblock \bibinfo{journal}{\emph{CACM}} \bibinfo{volume}{56},
  \bibinfo{number}{5} (\bibinfo{year}{2013}), \bibinfo{pages}{44–54}.
\newblock
\urldef\tempurl%
\url{https://doi.org/10.1145/2447976.2447990}
\showURL{%
\tempurl}


\bibitem[\protect\citeauthoryear{Telford}{Telford}{2019}]%
        {washPost}
\bibfield{author}{\bibinfo{person}{Taylor Telford}.}
  \bibinfo{year}{2019}\natexlab{}.
\newblock \bibinfo{title}{{Apple Card Algorithm Sparks Gender Bias Allegations
  against Goldman Sachs}}.
\newblock
  \bibinfo{howpublished}{\url{https://www.washingtonpost.com/business/2019/11/11/apple-card-algorithm-sparks-gender-bias-allegations-against-goldman-sachs/}}.
\newblock


\bibitem[\protect\citeauthoryear{Trimble and Kmec}{Trimble and Kmec}{2011}]%
        {trimble2011role}
\bibfield{author}{\bibinfo{person}{Lindsey~B Trimble} {and}
  \bibinfo{person}{Julie~A Kmec}.} \bibinfo{year}{2011}\natexlab{}.
\newblock \showarticletitle{The Role of Social Networks in Getting a Job}.
\newblock \bibinfo{journal}{\emph{Sociology Compass}} \bibinfo{volume}{5},
  \bibinfo{number}{2} (\bibinfo{year}{2011}), \bibinfo{pages}{165--178}.
\newblock


\bibitem[\protect\citeauthoryear{Vuolo, Uggen, and Lageson}{Vuolo
  et~al\mbox{.}}{2016}]%
        {vuolo2016statistical}
\bibfield{author}{\bibinfo{person}{Mike Vuolo}, \bibinfo{person}{Christopher
  Uggen}, {and} \bibinfo{person}{Sarah Lageson}.}
  \bibinfo{year}{2016}\natexlab{}.
\newblock \showarticletitle{Statistical Power in Experimental Audit Studies:
  Cautions and Calculations for Matched Tests with Nominal Outcomes}.
\newblock \bibinfo{journal}{\emph{Sociological Methods \& Research}}
  \bibinfo{volume}{45}, \bibinfo{number}{2} (\bibinfo{year}{2016}),
  \bibinfo{pages}{260--303}.
\newblock


\bibitem[\protect\citeauthoryear{Vuolo, Uggen, and Lageson}{Vuolo
  et~al\mbox{.}}{2018}]%
        {vuolo2018match}
\bibfield{author}{\bibinfo{person}{Mike Vuolo}, \bibinfo{person}{Christopher
  Uggen}, {and} \bibinfo{person}{Sarah Lageson}.}
  \bibinfo{year}{2018}\natexlab{}.
\newblock \showarticletitle{To Match or Not to Match? Statistical and
  Substantive Considerations in Audit Design and Analysis}.
\newblock In \bibinfo{booktitle}{\emph{Audit Studies: Behind the Scenes with
  Theory, Method, and Nuance}}. \bibinfo{publisher}{Springer},
  \bibinfo{address}{New York, NY}, \bibinfo{pages}{119--140}.
\newblock


\bibitem[\protect\citeauthoryear{Waldman}{Waldman}{2020}]%
        {waldman2020privacy}
\bibfield{author}{\bibinfo{person}{Ari~Ezra Waldman}.}
  \bibinfo{year}{2020}\natexlab{}.
\newblock \showarticletitle{Privacy Law’s False Promise}.
\newblock \bibinfo{journal}{\emph{Washington University Law Review}}
  \bibinfo{volume}{97}, \bibinfo{number}{3} (\bibinfo{year}{2020}),
  \bibinfo{pages}{773--834}.
\newblock


\bibitem[\protect\citeauthoryear{Whyte}{Whyte}{1989}]%
        {whyte1989advancing}
\bibfield{author}{\bibinfo{person}{William~F Whyte}.}
  \bibinfo{year}{1989}\natexlab{}.
\newblock \showarticletitle{Advancing Scientific Knowledge through
  Participatory Action Research}.
\newblock \bibinfo{journal}{\emph{Sociological Forum}} \bibinfo{volume}{4},
  \bibinfo{number}{3} (\bibinfo{year}{1989}), \bibinfo{pages}{367--385}.
\newblock


\bibitem[\protect\citeauthoryear{Wilson, Ghosh, Jiang, Mislove, Baker, Szary,
  Trindel, and Polli}{Wilson et~al\mbox{.}}{2021}]%
        {wilson2021building}
\bibfield{author}{\bibinfo{person}{Christo Wilson}, \bibinfo{person}{Avijit
  Ghosh}, \bibinfo{person}{Shan Jiang}, \bibinfo{person}{Alan Mislove},
  \bibinfo{person}{Lewis Baker}, \bibinfo{person}{Janelle Szary},
  \bibinfo{person}{Kelly Trindel}, {and} \bibinfo{person}{Frida Polli}.}
  \bibinfo{year}{2021}\natexlab{}.
\newblock \showarticletitle{Building and Auditing Fair Algorithms: A Case Study
  in Candidate Screening}. In \bibinfo{booktitle}{\emph{Proceedings of the 2021
  ACM Conference on Fairness, Accountability, and Transparency}}.
  \bibinfo{publisher}{ACM}, \bibinfo{address}{New York, NY},
  \bibinfo{pages}{666--677}.
\newblock


\bibitem[\protect\citeauthoryear{Wormser}{Wormser}{1950}]%
        {wormser1950conduct}
\bibfield{author}{\bibinfo{person}{Margot~Haas Wormser}.}
  \bibinfo{year}{1950}\natexlab{}.
\newblock \bibinfo{booktitle}{\emph{How to Conduct a Community Self-Survey of
  Civil Rights}}.
\newblock \bibinfo{publisher}{Association Press}, \bibinfo{address}{New York,
  NY}.
\newblock


\bibitem[\protect\citeauthoryear{Yee and Peradejordi}{Yee and
  Peradejordi}{2021}]%
        {TwitterBountyInsights}
\bibfield{author}{\bibinfo{person}{Kyra Yee} {and} \bibinfo{person}{Irene~Font
  Peradejordi}.} \bibinfo{year}{2021}\natexlab{}.
\newblock \bibinfo{title}{Sharing Learnings from the First Algorithmic Bias
  Bounty Challenge}.
\newblock
  \bibinfo{howpublished}{\url{https://blog.twitter.com/engineering/en_us/topics/insights/2021/learnings-from-the-first-algorithmic-bias-bounty-challenge}}.
\newblock


\bibitem[\protect\citeauthoryear{Yee, Tantipongpipat, and Mishra}{Yee
  et~al\mbox{.}}{2021}]%
        {yee2021image}
\bibfield{author}{\bibinfo{person}{Kyra Yee}, \bibinfo{person}{Uthaipon
  Tantipongpipat}, {and} \bibinfo{person}{Shubhanshu Mishra}.}
  \bibinfo{year}{2021}\natexlab{}.
\newblock \bibinfo{title}{Image Cropping on Twitter: Fairness Metrics, their
  Limitations, and the Importance of Representation, Design, and Agency}.
\newblock
\newblock
\showeprint{arXiv:2105.08667}


\bibitem[\protect\citeauthoryear{Yinger}{Yinger}{1998}]%
        {yinger1998evidence}
\bibfield{author}{\bibinfo{person}{John Yinger}.}
  \bibinfo{year}{1998}\natexlab{}.
\newblock \showarticletitle{Evidence on Discrimination in Consumer Markets}.
\newblock \bibinfo{journal}{\emph{Journal of Economic Perspectives}}
  \bibinfo{volume}{12}, \bibinfo{number}{2} (\bibinfo{year}{1998}),
  \bibinfo{pages}{23--40}.
\newblock


\bibitem[\protect\citeauthoryear{Zuberi and Bonilla-Silva}{Zuberi and
  Bonilla-Silva}{2008a}]%
        {bonilla2008toward}
\bibfield{author}{\bibinfo{person}{Tukufu Zuberi} {and}
  \bibinfo{person}{Eduardo Bonilla-Silva}.} \bibinfo{year}{2008}\natexlab{a}.
\newblock \bibinfo{booktitle}{\emph{White Logic, White Methods: Racism and
  Methodology}}.
\newblock \bibinfo{publisher}{Rowman \& Littlefield Publishers},
  \bibinfo{address}{Lanham, MD}, Chapter~1, \bibinfo{pages}{3--30}.
\newblock


\bibitem[\protect\citeauthoryear{Zuberi and Bonilla-Silva}{Zuberi and
  Bonilla-Silva}{2008b}]%
        {zuberi2008white}
\bibfield{author}{\bibinfo{person}{Tukufu Zuberi} {and}
  \bibinfo{person}{Eduardo Bonilla-Silva}.} \bibinfo{year}{2008}\natexlab{b}.
\newblock \bibinfo{booktitle}{\emph{White Logic, White Methods: Racism and
  Methodology}}.
\newblock \bibinfo{publisher}{Rowman \& Littlefield Publishers},
  \bibinfo{address}{Lanham, MD}.
\newblock


\end{thebibliography}

\end{document}